\documentclass [11pt,a4paper]{article}
\usepackage{jcappub}

\usepackage{amsmath}
\usepackage{amsfonts}
\usepackage{graphicx}
\usepackage{url}
\usepackage{bm}

\newcommand{\C}{{\cal C}}
\newcommand{\s}{{\cal S}}

\newcommand{\U}{{\cal U}}
\newcommand{\A}{{\cal A}}

\newcommand{\G}{{\cal G}}
\newcommand{\rs}{{\rm s}}
\newcommand{\rt}{{\rm t}}

\def\be{\begin{equation}}
\def\ee{\end{equation}}
\def\ba{\begin{eqnarray}}
\def\ea{\end{eqnarray}}
\def\bs{\begin{subequations}}
\def\es{\end{subequations}}
\def\com{\color{magenta}}
\def\cob{\color{blue}}

\newcommand{\arX}[1]{\href{http://arxiv.org/abs/#1}{{\ttfamily\com arXiv:#1}}}
\newcommand{\doij}[5]{\href{http://dx.doi.org/#1}{\cob {\it #2} {\bf #3} (#5) #4}}

\begin{document}

\title{Disformal invariance of cosmological perturbations 
in a generalized class of Horndeski theories} 

\author{Shinji Tsujikawa} 
\affiliation{Department of Physics, Faculty of Science, Tokyo University of Science, 
1-3, Kagurazaka, Shinjuku-ku, Tokyo 162-8601, Japan}
\emailAdd{shinji@rs.kagu.tus.ac.jp}

\abstract{It is known that Horndeski theories can be transformed to 
a sub-class of Gleyzes-Langlois-Piazza-Vernizzi (GLPV) 
theories under the disformal transformation of the metric 
$g_{\mu \nu} \to \Omega^2(\phi)g_{\mu \nu}+\Gamma (\phi,X)
\nabla_{\mu} \phi \nabla_{\nu} \phi$, where $\Omega$ is a function 
of a scalar field $\phi$ and $\Gamma$ is another function depending 
on both $\phi$ and $X=g^{\mu \nu}\nabla_{\mu} \phi \nabla_{\nu} \phi$. 
We show that, with the choice of unitary gauge,  
both curvature and tensor perturbations on the flat isotropic 
cosmological background are generally invariant under 
the disformal transformation. By means of the effective field theories 
encompassing Horndeski and GLPV theories,  
we obtain the second-order actions of scalar/tensor perturbations 
and present the relations for physical quantities between 
the two frames. The invariance of 
the inflationary power spectra under the disformal transformation 
is explicitly proved up to next-to-leading order in slow-roll. 
In particular, we identify the existence of the Einstein frame in 
which the tensor power spectrum is of the same form as that 
in General Relativity and derive the condition under which the spectrum of 
gravitational waves in GLPV theories is red-tilted. 
}
 


\maketitle

\section{Introduction}

The observational evidence of inflation and dark energy  
has pushed forward the idea that some scalar degree of freedom 
beyond the realms of General Relativity (GR) and the standard 
model of particle physics may be responsible for the two phases 
of cosmic accelerations \cite{review}. 
One of the well known modified gravitational theories
is Brans-Dicke (BD) theory \cite{Brans}, 
in which a scalar field $\phi$ couples to 
the Ricci scalar $R$ in the form $\phi R$. 
If we allow the presence of a scalar-field potential, 
BD theory can accommodate the metric $f(R)$ gravity as a specific case 
(the BD parameter $\omega_{\rm BD}=0$ \cite{Ohanlon,fRreview}).
The Starobinsky model $f(R)=R+R^2/(6M^2)$ gives rise to inflation 
due to the dominance of the $R^2$ term \cite{Staro}. 
We also have dark energy models constructed in the framework 
of $f(R)$ gravity \cite{fR1,fR2} and BD theory \cite{BDpapers}.

There are other modified gravitational theories like 
Galileons \cite{Nicolis,Galileons} in which the field kinetic term 
$X=g^{\mu \nu}\nabla_{\mu}\phi \nabla_{\nu}\phi$ couples to 
the Ricci scalar and the Einstein tensor (where $\nabla_{\mu}$ 
represents a covariant derivative). 
In such theories the cosmic acceleration can be driven by the 
field kinetic energy even without a scalar potential \cite{Gali1,Gali2,Gali3}. 
The Lagrangian of covariant Galileons is constructed to keep 
the equations of motion up to second order, while recovering the 
Galilean symmetry $\nabla_{\mu} \phi \to \nabla_{\mu} \phi+b_{\mu}$ 
in the limit of Minkowski space-time \cite{Galileons}. 
Without imposing the Galilean symmetry, it is possible to obtain
the Lagrangian of most general scalar-tensor theories 
with second-order equations of motion in generic 
space-time \cite{Deffayet,KYY,Char}. 
In fact, this was first derived by Horndeski in 1973 \cite{Horndeski}.

Horndeski theories encompass a wide variety of gravitational 
theories including BD theory. 
In BD theory the conformal transformation of the metric
$g_{\mu \nu} \to \Omega^2 (\phi)g_{\mu \nu}$, where $\Omega$ 
is a function of $\phi$, can give rise to a metric frame (dubbed 
Einstein frame) in which the field $\phi$ does not have a direct 
coupling to the Ricci scalar $\hat{R}$ \cite{Fujii,Maeda,Faraoni}.
The situation is more involved in Horndeski theories, but 
it was shown in Refs.~\cite{Liberati,Garcia} that the so-called 
disformal transformation in the form 
$g_{\mu \nu} \to \Omega^2 (\phi) g_{\mu \nu}
+\Gamma (\phi) \nabla_{\mu} \phi 
\nabla_{\nu} \phi$ \cite{Beken}
preserves the structure of the original Horndeski action.
Then, it should be possible to identify the Einstein frame in which 
the field does not have a direct coupling with $\hat{R}$.

Recently, Gleyzes, Langlois, Piazza, and Vernizzi (GLPV) \cite{Gleyzes} 
proposed a generalized class of Horndeski theories 
with second-order equations of motion on the flat 
Friedmann-Lema\^{i}tre-Robertson-Walker (FLRW) background.
According to the Hamiltonian analysis based on linear 
cosmological perturbations, GLPV theories have one 
scalar degree of freedom without ghost-like Ostrogradski 
instabilities \cite{Gleyzes,Lin,GlHa,GaoHa}. 
In Ref.~\cite{GlHa} it was shown that the structure 
of the GLPV action in unitary gauge is preserved 
under the disformal transformation 
$g_{\mu \nu} \to \Omega^2 (\phi) g_{\mu \nu}
+\Gamma (\phi,X) \nabla_{\mu} \phi \nabla_{\nu} \phi$. 
Hence the dependence of the function $\Gamma$ on $X$ 
generates terms absent in Horndeski theories.

The disformal transformation $g_{\mu \nu} \to \Omega^2 (\phi) g_{\mu \nu}
+\Gamma (\phi,X) \nabla_{\mu} \phi \nabla_{\nu} \phi$ is very useful to 
understand the relation between Horndeski and GLPV theories. 
As we will see in Sec.~\ref{GLPVsec}, the disformal transformation 
of Horndeski theories gives rise to a sub-class of GLPV theories satisfying one 
additional condition. Conversely, the transformation from GLPV theories 
to Horndeski theories demands that the factor $\Gamma$ obeys 
two different conditions simultaneously \cite{GlHa}.
Thus the full Horndeski and GLPV theories are not equivalent to each other, 
but the two non-Horndeski pieces in the GLPV action separately arise from 
a subset of the Horndeski action under the disformal transformation.

In BD and non-minimally coupled theories, it is known that 
both scalar and tensor perturbations are invariant 
under the conformal 
transformation \cite{Fakir,Makino,Kaiser,Hwang,Gong,White,Chiba2} 
(see also Refs.~\cite{Catena,Nata}).
In fact, this equivalence was used for the computation of 
the spectral indices of the primordial power spectra and the tensor-to-scalar 
ratio to confront non-minimally coupled inflationary models
with observations \cite{Komatsu,Gum,Kuro} 
(e.g., Higgs inflation \cite{FM,Higgs}).
The invariance of cosmological perturbations under the 
disformal transformation $g_{\mu \nu} \to \Omega^2 (\phi) g_{\mu \nu}
+\Gamma (\phi) \nabla_{\mu} \phi \nabla_{\nu} \phi$ in 
Horndeski theories was recently proved in Ref.~\cite{Minami}.
In this paper, we show the frame independence of 
curvature and tensor perturbations on the flat FLRW background
under the more general disformal transformation 
$g_{\mu \nu} \to \Omega^2 (\phi) g_{\mu \nu}
+\Gamma (\phi,X) \nabla_{\mu} \phi \nabla_{\nu} \phi$.

By means of the effective field theory (EFT) of cosmological 
perturbations developed in Refs.~\cite{Cheung,Weinberg,Quin,Park,Bloom,Gubi,Piazza}, 
we obtain the second-order actions of scalar and tensor perturbations by 
choosing the unitary gauge. This procedure is closely related to the analysis 
in Refs.~\cite{Piazza,Gergely,Tsuji14}, but the background lapse dependence is 
explicitly taken into account for the quantities associated with linear perturbations.
The latter treatment is important for understanding the relation between 
the quantities in the two frames connected by the disformal transformation.

In the EFT approach we also derive the primordial power spectra of scalar 
and tensor perturbations generated during inflation up to next-to-leading order in slow-roll. 
The invariance of inflationary observables (such as the spectral 
indices and the tensor-to-scalar ratio) under the disformal transformation 
is explicitly shown by paying particular attention to the change of quantities 
associated with the perturbation equations of motion.
Moreover, we show the existence of the Einstein frame 
in which the next-to-leading order tensor power spectrum is 
of the same form as that in GR. We also study the background equations of motion 
in the Einstein frame and derive the condition under which the 
inflationary tensor power spectrum is red-tilted in GLPV theories.

This paper is organized as follows.
In Sec.~\ref{disformalsec} we show the general invariance of 
curvature and tensor perturbations under the disformal 
transformation $g_{\mu \nu} \to \Omega^2 (\phi) g_{\mu \nu}
+\Gamma (\phi,X) \nabla_{\mu} \phi \nabla_{\nu} \phi$ in unitary gauge.
In Sec.~\ref{GLPVsec} we discuss how the structure of the GLPV 
action is preserved under the disformal transformation.
In Sec.~\ref{persec} the second-order actions of scalar 
and tensor perturbations are derived in the EFT approach 
encompassing both Horndeski and GLPV theories.
In Sec.~\ref{corressec} we present explicit relations between 
the quantities associated with the background and perturbations 
in the two frames linked through the disformal transformation by considering GLPV theories. 
In Sec.~\ref{powerspssec} we apply the results in Sec.~\ref{persec} 
to the derivation of the inflationary power spectra 
up to next-to-leading order in slow-roll
and show their invariance under the disformal transformation. 
In Sec.~\ref{einsteinsec} we identity the existence of the Einstein frame 
by appropriately choosing the functions $\Omega$ and $\Gamma$.
Sec.~\ref{consec} is devoted to conclusions.

\section{Disformal transformation} \label{disformalsec}

We begin with the line element based on the 
Arnowitt-Deser-Misner (ADM) formalism \cite{ADM} given by 
\be
ds^2=g_{\mu \nu} dx^{\mu} dx^{\nu}
=-N^2 dt^2+h_{ij} (dx^i+N^i dt) 
(dx^j+N^j dt)\,,
\label{Admmetric}
\ee
where $N$ is the lapse, $N^i$ is the shift vector, 
$g_{\mu \nu}$ and $h_{ij}$ are the four-dimensional 
and three-dimensional metrics respectively. 
Throughout the paper, Greek and Latin indices represent 
components in space-time and 
in a three-dimensional space-adapted basis, respectively.
The perturbed line element on the flat FLRW 
background is characterized by \cite{Bardeen}
\be
ds^2=-(1+2A)dt^2+2 \psi_{|i} dt dx^i
+a^2(t) \left[ (1+2\zeta) \delta_{ij}+\gamma_{ij} 
+2E_{|ij} \right]dx^i dx^j\,,
\label{permet}
\ee
where $a(t)$ is the scale factor that depends on 
the cosmic time $t$, the lower index ``${}_{|i}$'' denotes the covariant 
derivative with respect to the three-dimensional metric 
$h_{ij}$, and $A, \psi, \zeta, E$ are the scalar metric perturbations 
and $\gamma_{ij}$ is the tensor perturbation.
Comparing Eq.~(\ref{Admmetric}) with Eq.~(\ref{permet}), 
there is the correspondence 
$1+2A=N^2-h_{ij}N^i N^j$, $\psi_{|i}=h_{ij}N^j$, and 
\be
h_{ij}=a^2(t) q_{ij}\,,\quad {\rm where} \quad
q_{ij} \equiv (1+2\zeta) \delta_{ij}+\gamma_{ij} +2E_{|ij}\,.
\label{hij}
\ee

For the line element (\ref{Admmetric}) we perform the following 
disformal transformation  
\be
\hat{g}_{\mu \nu}=\Omega^2 (\phi) g_{\mu \nu}
+\Gamma (\phi,X) \nabla_{\mu} \phi 
\nabla_{\nu} \phi\,,
\label{distra}
\ee
where $\Omega (\phi)$ is a function of
a scalar field $\phi$, and $\Gamma (\phi,X)$ is 
a function that depends on $\phi$ and its kinetic 
energy $X=g^{\mu \nu} \nabla_{\mu}\phi \nabla_{\nu}\phi$. 
In the following we use a hat for the quantities in the 
transformed frame.
We choose the unitary gauge in which 
$\phi$ depends on the time $t$ alone, i.e., 
\be
\phi=\phi(t)\,.
\label{unitary}
\ee
In this case the field kinetic energy is given by 
$X=-N^{-2} \dot{\phi}^2$, where a dot represents 
a derivative wit respect to $t$.
Hence the dependence on $\phi$ and $X$ in $\Gamma$ 
can be interpreted as that on $t$ and $N$. 
The line element in the transformed 
frame reads
\ba
d\hat{s}^2=\hat{g}_{\mu \nu} dx^{\mu} dx^{\nu}
&=&
-N^2( \Omega^2 +\Gamma X )
dt^2+\Omega^2 h_{ij} (dx^i+N^i dt)(dx^j +N^j dt) \nonumber \\
&=&-\hat{N}^2 dt^2+\hat{h}_{ij}
(dx^i+N^i dt)(dx^j +N^j dt)\,,
\label{trans1}
\ea
where in the second line $\hat{N}$ and $\hat{h}_{ij}$ are
given, respectively, by 
\ba
\hat{N} &=&N\sqrt{\Omega^2+\Gamma X}\,,
\label{hatN} \\
\hat{h}_{ij} &=& \Omega^2 h_{ij}\,.
\label{hijtra}
\ea

In unitary gauge the conformal factor $\Omega (\phi)$ depends on 
$t$ but not on $x^i$, so we can introduce the scale factor
in the transformed frame:
\be
\hat{a}(t)=\Omega a(t)\,.
\label{hata}
\ee
Using Eqs.~(\ref{hij}), (\ref{hijtra}), and (\ref{hata}),
the line element (\ref{trans1}) can be expressed as
\be
d\hat{s}^2=-\hat{N}^2 dt^2+\hat{a}^2(t) 
q_{ij} (dx^i+N^i dt)(dx^j +N^j dt)\,.
\label{trans2}
\ee
Then the three-dimensional tensor $q_{ij}$ and 
the shift $N^i$ are invariant under the 
disformal transformation, such that 
\ba
\hat{\zeta} &=& \zeta\,,\label{zeta1} \\
\hat{\gamma}_{ij} &=& \gamma_{ij}\,,\label{gamma1}\\
\hat{N}^i &=& N^i\,, \label{Ni}
\ea
and $\hat{E}=E$.
In unitary gauge (\ref{unitary}) where the field perturbation 
$\delta \phi$ vanishes, the scalar perturbation $\zeta$ 
itself is a gauge-invariant quantity \cite{Bran}. 
Thus, from Eqs.~(\ref{zeta1}) and (\ref{gamma1}),
both the curvature perturbation $\zeta$ and the tensor 
perturbation $\gamma_{ij}$ are invariant under the disformal 
transformation (\ref{distra}). 
This is the generalization of the results of Ref.~\cite{Minami}
in which the same disformal invariance was shown for the 
transformation $\hat{g}_{\mu \nu}=\Omega^2 (\phi) g_{\mu \nu}
+\Gamma (\phi) \nabla_{\mu} \phi 
\nabla_{\nu} \phi$.

While we have shown the invariance of $\zeta$ and $\gamma_{ij}$ 
under the disformal transformation, it remains to see the transformation 
properties for quantities appearing in the background and 
perturbation equations of motion. In GLPV theories we shall 
address this problem in Secs.~\ref{GLPVsec} and \ref{corressec}. 
Derivation of the relations for quantities in the two different frames
is particularly important to identify the tensor and scalar power spectra 
in the Einstein frame. This issue is addressed in Sec.~\ref{einsteinsec}.

\section{Disformal transformation in GLPV theories} \label{GLPVsec}

The four-dimensional Lagrangian of the most general scalar-tensor 
theories with second-order equations of motion 
(Horndeski theories \cite{Horndeski}) 
is given by the action \cite{Deffayet,KYY}
\be
S=\int d^4 x \sqrt{-g}\,L\,,
\label{oriac}
\ee
where $g$ is the determinant of the metric $g_{\mu \nu}$, and 
\ba
L &=& G_2(\phi,X)+G_{3}(\phi,X)\square\phi 
+G_{4}(\phi,X)\, R-2G_{4,X}(\phi,X)\left[ (\square \phi)^{2}
-\phi^{;\mu \nu }\phi _{;\mu \nu} \right] \nonumber \\
& &
+G_{5}(\phi,X)G_{\mu \nu }\phi ^{;\mu \nu}
+\frac{1}{3}G_{5,X}(\phi,X)
[ (\square \phi )^{3}-3(\square \phi )\,\phi _{;\mu \nu }\phi ^{;\mu
\nu }+2\phi _{;\mu \nu }\phi ^{;\mu \sigma }{\phi ^{;\nu}}_{;\sigma}]\,,
\label{Lho}
\ea
where a semicolon represents a covariant derivative with
$\square \phi \equiv (g^{\mu \nu} \phi_{;\nu})_{;\mu}$, 
$R$ is the Ricci scalar, and  $G_{\mu\nu}$ is 
the Einstein tensor. The four functions $G_i$ ($i=2,3,4,5$) 
depend on $\phi$ and $X$ with the partial derivatives 
$G_{i,X} \equiv \partial G_i/\partial X$ and 
$G_{i,\phi} \equiv \partial G_i/\partial \phi$.

The Horndeski Lagrangian (\ref{Lho}) can be reformulated 
by using geometric scalar quantities appearing in the 
ADM formalism \cite{Piazza}. Defining the extrinsic curvature  
as $K_{\mu \nu}=h^{\lambda}_{\mu} n_{\nu;\lambda}$, where 
$n_{\mu}=(-N,0,0,0)$ is a unit vector orthogonal to 
the constant $t$ hypersurfaces $\Sigma_t$, we can 
construct the following scalar quantities
\be
K \equiv {K^{\mu}}_{\mu}\,,\qquad
{\cal S} \equiv K_{\mu \nu} K^{\mu \nu}\,.
\ee
The three-dimensional Ricci tensor 
${\cal R}_{\mu \nu}={}^{(3)}R_{\mu \nu}$ (intrinsic curvature) 
characterizes the internal geometry of $\Sigma_t$. 
The scalar quantities constructed from ${\cal R}_{\mu \nu}$ 
and $K_{\mu \nu}$ are given by 
\be
{\cal R} \equiv
{{\cal R}^{\mu}}_{\mu}\,,\qquad
{\cal U} \equiv \mathcal{R}_{\mu \nu}K^{\mu \nu}\,,\qquad
{\cal Z} \equiv {\cal R}_{\mu \nu}
\mathcal{R}^{\mu \nu}\,. \qquad
\ee

Choosing the unitary gauge on the flat FLRW background, 
the dependence on $\phi$ and $X$ in 
the functions $G_i$ can be interpreted as that on 
$t$ and $N$. Expressing the scalar quantities like 
$\square \phi$ and $R$ in terms of the three-dimensional 
geometric scalars mentioned above, the Horndeski 
Lagrangian (\ref{Lho}) is equivalent to \cite{Piazza,Gleyzes,GLV}
\ba
L &=&
A_2(N,t)+A_3(N,t)K+A_4(N,t) (K^2-{\cal S})
+B_4(N,t){\cal R} \nonumber \\
&&+A_5(N,t) K_3
+B_5(N,t) \left( {\cal U}-K {\cal R}/2 \right)\,,
\label{LH}
\ea
where 
\be
K_3=K^3-3K{\cal S}
+2K_{\mu \nu}K^{\mu \lambda}{K^{\nu}}_{\lambda}\,,
\label{K3}
\ee
and 
\ba
& & A_2=G_2-XF_{3,\phi}\,,\qquad
A_3=2(-X)^{3/2}F_{3,X}-2\sqrt{-X}G_{4,\phi}\,,\label{A3}
\nonumber \\
& & A_4=-G_4+2XG_{4,X}+XG_{5,\phi}/2\,,\qquad
B_4=G_4+X(G_{5,\phi}-F_{5,\phi})/2\,,\label{B4} 
\nonumber \\
& & A_5=-(-X)^{3/2}G_{5,X}/3\,,\qquad 
B_5=-\sqrt{-X}F_{5}\,.\label{B5}
\label{AB}
\ea
Here, $F_3 (\phi,X)$ and $F_5 (\phi,X)$ are auxiliary functions satisfying
$G_3=F_3+2XF_{3,X}$ and $G_{5,X}=F_5/(2X)+F_{5,X}$.
{}From Eq.~(\ref{AB}) the following two relations hold
\be
A_4=2XB_{4,X}-B_4\,,\qquad
A_5=-XB_{5,X}/3\,.
\label{ABcon}
\ee

The GLPV theories correspond to the Lagrangian (\ref{LH}) 
without the particular relations (\ref{ABcon}).
On the flat FLRW background the function $K_3$ 
can be expressed in terms of 
$K$ and ${\cal S}$ \cite{Piazza}, so 
the GLPV Lagrangian depends on 
$N, t, K, {\cal S}, {\cal R}, {\cal U}$ but not on ${\cal Z}$.
The dependence on ${\cal Z}$ arises for the theories 
with spatial derivatives higher than second 
order \cite{Xian,Kase,DeTsuji14}, e.g., 
Ho\v{r}ava-Lifshitz gravity \cite{Horava}.

Under the disformal transformation (\ref{distra}) the lapse 
$\hat{N}$ and the three-dimensional metric $\hat{h}_{ij}$ 
in the line element (\ref{trans1}) are given by 
Eqs.~(\ref{hatN}) and (\ref{hijtra}) respectively, 
so the volume element is transformed as \cite{GlHa}
\be
\sqrt{-\hat{g}}=\sqrt{-g}\,\Omega^3 \alpha\,,
\label{gtra}
\ee
where 
\be
\alpha \equiv \frac{\hat{N}}{N}=
\sqrt{\Omega^2+\Gamma X}\,.
\ee

In unitary gauge (\ref{unitary}), the conformal factor $\Omega (\phi)$ 
depends on $t$ but not on $N$.
The extrinsic curvature in the transformed frame is given by 
$\hat{K}_{ij}=(\partial \hat{h}_{ij}/\partial t-\hat{N}_{i|j}-\hat{N}_{j|i})/(2\hat{N})$.
On using Eqs.~(\ref{hijtra}) and (\ref{Ni}), it follows that 
$\hat{N}_j=\Omega^2(t)N_j$. Then the transformation 
of the extrinsic curvature reads
\be
\hat{K}_{ij}=\frac{\Omega^2}{\alpha} 
\left( K_{ij}+\frac{\omega}{N}h_{ij} \right)\,,
\label{Kijtra}
\ee
where 
\be
\omega \equiv \frac{\dot{\Omega}}{\Omega}\,.
\ee
The transformation (\ref{hijtra}) of the metric $h_{ij}$ 
is the same as the conformal transformation 
in three dimensions. 
Hence the three-dimensional Ricci tensor 
transforms as \cite{Wald}
\be
\hat{{\cal R}}_{ij}={\cal R}_{ij}-\nabla_i \nabla_j \ln \Omega
-g_{ij} g^{kl} \nabla_k \nabla_l \ln \Omega
+(\nabla_i \ln \Omega)(\nabla_j \ln \Omega)
-g_{ij}g^{kl}(\nabla_k \ln \Omega)(\nabla_l \ln \Omega)\,.
\label{Rijtrans}
\ee
Since $\Omega$ is a function of $t$ alone in unitary gauge, 
the spatial derivatives of $\Omega$ vanish in Eq.~(\ref{Rijtrans}). 
Then the transformations of ${\cal R}_{ij}$ and ${\cal R}$ 
are simply given by 
\be
\hat{{\cal R}}_{ij}={\cal R}_{ij}\,,\qquad
\hat{{\cal R}}=\Omega^{-2}{\cal R}\,.
\label{Rtra}
\ee

Employing the transformation laws (\ref{gtra}), (\ref{Kijtra}), and (\ref{Rtra}) 
for the action (\ref{oriac}) with the GLPV Lagrangian (\ref{LH}), 
the action in the transformed frame reads
\be
S=\int d^4 x \sqrt{-\hat{g}}\,\hat{L}\,,
\label{Stra}
\ee
where 
\ba
\hat{L} &=&
\hat{A}_2(\hat{N},t)+\hat{A}_3(\hat{N},t)\hat{K}
+\hat{A}_4(\hat{N},t) (\hat{K}^2-\hat{\cal S})
+\hat{B}_4(\hat{N},t)\hat{{\cal R}} \nonumber \\
&&+\hat{A}_5(\hat{N},t) \hat{K}_3
+\hat{B}_5(\hat{N},t) \left( \hat{\cal U}-
\hat{K} \hat{\cal R}/2 \right)\,,
\label{LH2}
\ea
with the coefficients \cite{GlHa}
\ba
\hat{A}_2 &=&
\frac{1}{\Omega^3 \alpha} \left( A_2-\frac{3\omega}{N}A_3
+\frac{6\omega^2}{N^2}A_4-\frac{6\omega^3}{N^3}A_5 \right)\,,
\label{hatA2}\\
\hat{A}_3 &=&
\frac{1}{\Omega^3}  \left( A_3-\frac{4\omega}{N}A_4
+\frac{6\omega^2}{N^2}A_5 \right)\,,
\label{hatA3}\\
\hat{A}_4 &=&
\frac{\alpha}{\Omega^3} \left( A_4-\frac{3\omega}{N}A_5 \right)\,,
\label{hatA4} \\
\hat{B}_4 &=&
\frac{1}{\Omega \alpha} \left( B_4+\frac{\omega}{2N}B_5 \right)\,,
\label{hatB4}\\
\hat{A}_5 &=&
\frac{\alpha^2}{\Omega^3}A_5\,,
\label{hatA5} \\
\hat{B}_5 &=&
\frac{1}{\Omega}B_5\,.
\label{hatB5}
\label{cotra}
\ea
The structure of the Lagrangian (\ref{LH2}) is the same as 
the original GLPV Lagrangian (\ref{LH}), so the disformal 
transformation (\ref{distra}) allows the connection between the GLPV theories.

Let us consider Horndeski theories described by the Lagrangian 
(\ref{LH}) satisfying the two conditions (\ref{ABcon}). 
On using Eqs.~(\ref{hatA4})-(\ref{hatB5}) and the correspondence
$\hat{X}=\alpha^{-2} X$, we obtain 
\ba
& &
\hat{A}_4+\hat{B}_4-2\hat{X}\hat{B}_{4,\hat{X}}
=-\frac{X^2 \Gamma_{,X}}{\Omega^2-X^2 \Gamma_{,X}} \hat{A}_4\,,
\label{Hore1} \\
& &
\hat{A}_5+\frac13 \hat{X} \hat{B}_{5,\hat{X}}=
-\frac{X^2 \Gamma_{,X}}{\Omega^2-X^2 \Gamma_{,X}} \hat{A}_5\,.
\label{Hore2}
\ea
If $\Gamma$ is a function of $\phi$ alone, it follows that 
$\hat{A}_4+\hat{B}_4-2\hat{X}\hat{B}_{4,\hat{X}}=0$ 
and $\hat{A}_5+\hat{X} \hat{B}_{5,\hat{X}}/3=0$.
Hence, as shown in Refs.~\cite{Liberati,Garcia}, 
the disformal transformation 
of the form $\hat{g}_{\mu \nu}=\Omega^2 (\phi)g_{\mu \nu}
+\Gamma (\phi) \nabla_{\mu} \phi \nabla_{\nu} \phi$ preserves the 
structure of the Horndeski action.
If $\Gamma$ depends on both $\phi$ and $X$, 
Horndeski theories are transformed to a sub-class of 
GLPV theories obeying the particular relation 
\be
\frac{\hat{B}_4-2\hat{X}\hat{B}_{4,\hat{X}}}{\hat{A}_4}
=\frac{\hat{X}\hat{B}_{5,\hat{X}}}{3\hat{A}_5}\,, 
\ee
which follows from Eqs.~(\ref{Hore1}) and (\ref{Hore2}).
Conversely, the full action of GLPV theories cannot be generally 
mapped to  that in Horndeski theories because the function $\Gamma$  
needs to be chosen to satisfy the two Horndeski conditions 
simultaneously \cite{GlHa}.

\section{Second-order actions of cosmological perturbations} \label{persec}

In the EFT of modified gravity including both Horndeski and GLPV theories, 
the perturbation equations on the flat FLRW background were derived 
in Refs.~\cite{Piazza,Gergely}. 
In these papers the background value of the lapse $N$ (denoted as $\bar{N}$)
is set to 1 after obtaining the background and perturbation equations 
of motion. Since the lapse is transformed as Eq.~(\ref{hatN}) under 
the disformal transformation, we do not set $\bar{N}=1$ in the following discussion. 
Since the Lagrangian (\ref{LH}) in GLPV theories involves the dependence 
on $N, K, {\cal S}, {\cal R}, {\cal U}$ and $t$, 
we expand the following action 
\be
S=\int d^4 x \sqrt{-g}\,L(N,K, {\cal S}, {\cal R}, {\cal U};t)\,,
\label{geneac}
\ee
up to quadratic order in perturbations. 
For the partial derivatives of $L$ with respect to scalar quantities,
we use the notation like $L_{,N} \equiv \partial L/\partial N$ and 
$L_{,K} \equiv \partial L/\partial K$.

On the flat FLRW background described by the line element 
$ds^2=-\bar{N}^2dt^2+a^2(t) \delta_{ij}dx^i dx^j$, the 
geometric ADM quantities are given by  
\be
\bar{K}_{\mu \nu}=H\bar{h}_{\mu \nu}\,,
\qquad \bar{K}=3H\,,\qquad 
\bar{\cal S}=3H^{2}\,,\qquad \bar{{\cal R}}_{\mu \nu}=0\,,
\qquad \bar{{\cal R}}=\bar{\cal U}=0\,,
\ee
where a bar represents background quantities, and 
$H$ is the Hubble parameter defined by 
\be
H \equiv \frac{\dot{a}}{\bar{N}a}\,.
\ee
We also introduce the following perturbed quantities
\be
\delta N=N-\bar{N}\,,\qquad
\delta K_{\mu \nu}=K_{\mu \nu}-Hh_{\mu \nu}\,,
\qquad \delta K=K-3H\,,\qquad 
\delta {\cal S}=2H\delta K+\delta
K_{\nu }^{\mu} \delta K_{\mu}^{\nu}\,.
\ee
Since the intrinsic curvature ${\cal R}$ vanishes on the background, 
we can write
\be
{\cal R}=\delta_1 {\cal R}+\delta_2 {\cal R}\,,
\ee
where $\delta_1 {\cal R}$ and $\delta_2 {\cal R}$ are the first-order 
and second-order perturbations, respectively.
The scalar ${\cal U}$ is also a perturbed quantity, which satisfies 
the following relation (up to a boundary term) \cite{Piazza}
\be
\lambda(t) \,{\cal U}=\frac12 \lambda(t) {\cal R} K+
\frac{1}{2N} \dot{\lambda}(t){\cal R}\,,
\ee
where $\lambda(t)$ is an arbitrary function with respect to $t$.

Expanding the action (\ref{geneac}) up to first and second order 
in scalar perturbations, we can derive the background and 
scalar perturbation equations of motion respectively. 
In order to fix the temporal and spatial transformation vectors 
associated with coordinate transformations, we choose 
the unitary gauge
\be
\delta \phi=0\,, \qquad E=0\,,
\ee
where the former corresponds to Eq.~(\ref{unitary}).

\subsection{Background equations}

Following the same procedure as that given in Refs.~\cite{Piazza,Gergely}, 
the first-order action of scalar perturbations 
reduces to $S^{(1)}=\int d^4 x\,{\cal L}_1$ with 
\be
{\cal L}_1=a^3 \left( \bar{L}+\bar{N}L_{,N}-3H {\cal F} \right) \delta N
+\bar{N} \left( \bar{L}-\frac{\dot{\cal F}}{\bar{N}}-3H{\cal F} \right)
\delta \sqrt{h}+\bar{N}a^3 {\cal E} \delta_1 {\cal R}\,,
\label{L1}
\ee
where $h$ is the determinant of the three-dimensional 
metric $h_{ij}$, and
\ba
{\cal F} &\equiv& L_{,K}+2H L_{,{\cal S}}\,,\\
{\cal E} &\equiv& L_{,{\cal R}}+\frac{\dot{L}_{,\cal U}}{2\bar{N}}
+\frac32 HL_{,\cal U}\,.
\ea
The last term of Eq.~(\ref{L1}) is a total derivative irrelevant to 
the dynamics. Varying the action $S^{(1)}$ with respect to 
$\delta N$ and $\delta \sqrt{h}$, we obtain the background 
equations of motion 
\ba
& & \bar{L}+\bar{N}L_{,N}-3H {\cal F}=0\,,\label{back1}\\
& & \bar{L}-\frac{\dot{\cal F}}{\bar{N}}-3H{\cal F}=0\,,
\label{back2}
\ea
respectively.

\subsection{Second-order action of scalar perturbations}

Expanding the action (\ref{geneac}) up to quadratic order 
in scalar perturbations, we obtain the second-order action 
$S^{(2)}=\int d^4 x\,{\cal L}_2$ with 
\ba
\mathcal{L}_2 &=& a^3 \bar{N} \biggl[ \left\{
\frac{L_{,N}}{\bar{N}}+\frac12 L_{,NN}
-\frac{3H}{\bar{N}} \left( {\cal W}+
\frac{3{\cal A}H}{2\bar{N}}+\frac{L_{,\cal S}H}
{\bar{N}} \right) \right\} \delta N^2
\nonumber \\
& &~~~~~~
+\left\{ \frac{{\cal W}}{\bar{N}} ( 3\dot{\zeta}-\Delta \psi ) 
+\frac{4}{\bar{N}}(3H \C-\bar{N}{\cal D}-\mathcal{E}) 
\Delta \zeta \right\} \delta N 
-(3 \A+2L_{,\s}) \frac{\dot{\zeta}}{{\bar N}^2} \Delta \psi 
\nonumber \\
& &~~~~~~
-12\C \frac{\dot{\zeta}}{\bar N} \Delta \zeta+
\left( \frac92 {\cal A}+3L_{,\s} \right) \frac{\dot{\zeta}^2}{{\bar N}^2} 
+2\mathcal{E} 
\frac{(\partial \zeta)^2}{a^2} \nonumber \\
& &~~~~~~
+\frac12 ( \A+2L_{,\s} ) \frac{(\Delta \psi)^2}{{\bar N}^2}
+4\frac{\C}{\bar N} (\Delta \psi)(\Delta \zeta) 
+8\G (\Delta \zeta)^2 \biggr]\,,
\label{L2exp}
\ea
where $(\partial \zeta)^2=\delta^{ij}(\partial_i \zeta)(\partial_j \zeta)
=\delta^{ij} (\partial \zeta/\partial x^i)(\partial \zeta/\partial x^j)$, 
$\Delta=\nabla_i \nabla^i=a^{-2}(t)\delta^{ij} 
\partial_i \partial_j \equiv a^{-2}(t) \partial^2$, and  
\begin{eqnarray}
\A &=&L_{,KK}+4HL_{,K \s} +4H^{2}L_{,\s \s}\,,\\
\C &=&L_{,K\mathcal{R}}+2HL_{,\s \mathcal{R}}
+\frac{1}{2} L_{,\U}+HL_{,K \U}+2H^{2}L_{,\s \U}\,, \\
{\cal D} &=&L_{,N\mathcal{R}}-\frac{\dot{L_{,\U}}}{2\bar{N}^2}
+HL_{,N \U}\,, \\
\G &=&L_{,\mathcal{R}\mathcal{R}}
+2HL_{,\mathcal{R}\U}+H^{2}L_{,\U \U}\,,\\
{\cal W} &=& 
L_{,KN}+2HL_{,{\cal S}N}
-\frac{H}{\bar{N}} (3\A+2L_{,\s})\,.
\end{eqnarray}

On using the fact that the term $K_3$ in the Lagrangian (\ref{LH}) 
is given by
\be
K_3=3H(K^2-\s-2KH+2H^2)
\ee
up to quadratic order in perturbations \cite{Piazza}, 
the GLPV theories (\ref{LH}) obey the three conditions 
\be
\A+2L_{,\s}=0\,,\qquad {\cal C}=0\,,\qquad
{\cal G}=0\,,
\label{dericon}
\ee
under which the spatial derivatives higher than second order 
are absent in Eq.~(\ref{L2exp}). 
We shall employ the conditions (\ref{dericon}) in the 
following discussion.
Varying the action (\ref{L2exp}) with respect to $\delta N$ and 
$\psi$, we obtain the Hamiltonian and momentum 
constraints respectively:
\ba
\hspace{-0.9cm}& & \left( 2L_{,N}+\bar{N}L_{,NN}-6H {\cal W}
+\frac{12H^2L_{,\s}}{{\bar N}} \right) \delta N
+\left( 3\dot{\zeta}-\frac{\partial^2 \psi}{a^2} 
\right) {\cal W}-4({\bar N}{\cal D}+{\cal E}) 
\frac{\partial^2 \zeta}{a^2}=0\,,\label{scaper1} \\
\hspace{-0.9cm}& & {\cal W} \delta N
-4L_{,\s} \frac{\dot{\zeta}}{{\bar N}}=0\,.
\label{scaper2}
\ea
Expressing $\delta N$ and $\partial^2 \psi/a^2$ in terms $\dot{\zeta}$ 
and $\partial^2 \zeta/a^2$ from Eqs.~(\ref{scaper1})-(\ref{scaper2})
and substituting them into Eq.~(\ref{L2exp}), the second-order 
Lagrangian density (\ref{L2exp}) reduces to (up to boundary terms)
\be
{\cal L}_2=a^3 q_{\rs} \left[ \dot{\zeta}^2-\frac{c_{\rs}^2}{a^2} 
(\partial \zeta)^2 \right]\,,
\label{L2}
\ee
where 
\ba
q_{\rs} &=& \frac{2L_{,\s}[4L_{,\s}(2\bar{N}L_{,N}+\bar{N}^2L_{,NN})
+3(\bar{N}{\cal W}-4HL_{,\s})^2]}{\bar{N}^3 {\cal W}^2}\,,\label{qs} \\
c_{\rs}^2 &=& \frac{2\bar{N}}{q_\rs} \left( \frac{\dot{\cal M}}{\bar{N}}
+H{\cal M}-{\cal E} \right)\,,\label{cs}
\ea
and 
\be
{\cal M}=\frac{4L_{,\s}({\bar N}{\cal D}+{\cal E})}
{{\bar N}{\cal W}}
=\frac{4L_{,\s}}{{\bar N}{\cal W}}
\left( L_{,\cal R}+\bar{N}L_{,{\cal R}N}
+\frac32 H L_{,{\cal U}}+\bar{N}H
L_{,N{\cal U}}
\right)\,.
\ee
{}From the Lagrangian density (\ref{L2}) we obtain 
the equation of motion for scalar perturbations
\be
\frac{d}{dt} (a^3 q_{\rs} \dot{\zeta})
-aq_{\rs} c_{\rs}^2 \partial^2 \zeta=0\,,
\label{scaeq}
\ee
which is of second order. 
In order to avoid ghosts and Laplacian instabilities, we require that 
$q_{\rs}>0$ and $c_{\rs}^2>0$ respectively.
In Sec.~\ref{powerspssec}, we solve Eq.~(\ref{scaeq}) for 
the computation of the power spectrum of curvature 
perturbations generated during inflation.

\subsection{Second-order action of tensor perturbations}

We derive the second-order action $S_{2}^{(h)}=\int d^4 x \sqrt{-g}\,L_2^{(h)}$ 
for tensor perturbations $\gamma_{ij}$.
The non-vanishing terms in the second-order Lagrangian $L_2^{(h)}$ 
are $L_{,\s}\delta K^{i}_{j}\delta K^{j}_{i}$ and 
${\cal E}{\cal R}$, where $\delta K^{i}_{j}=\delta^{ik} \dot{\gamma}_{kj}/(2\bar{N})$ 
and ${\cal R}=\delta^{ik} \delta^{jl} \gamma_{ij} 
\Delta \gamma_{kl}$ \cite{Piazza,Tsuji14,DeTsuji14}.
Then, the second-order Lagrangian density 
${\cal L}_2^{(h)}=\sqrt{-g}L_2^{(h)}$ reads
\be
{\cal L}_2^{(h)}=a^3 q_{\rt} \delta^{ik} \delta^{jl}
\left( \dot{\gamma}_{ij} \dot{\gamma}_{kl}
-\frac{c_{\rt}^2}{a^2}\partial \gamma_{ij} 
\partial \gamma_{kl} \right)\,, 
\label{L2ten}
\ee
where 
\ba
q_{\rt} &=& \frac{L_{,\s}}{4\bar{N}}\,,\label{qtdef} \\
c_{\rt}^2 &=& \frac{\bar{N}^2 {\cal E}}{L_{,\s}}\,.
\label{ctdef}
\ea
The resulting equation of motion for tensor perturbations
is given by 
\be
\frac{d}{dt} (a^3 q_{\rt} \dot{\gamma}_{ij})
-aq_{\rt} c_{\rt}^2 \partial^2 \gamma_{ij}=0\,,
\label{tensoreq}
\ee
which is of the same form as Eq.~(\ref{scaeq}) apart from 
the difference of the coefficients $q_{\rt}$ and $c_{\rt}$.

\section{Relations between the two frames connected 
by the disformal transformation} \label{corressec}

In GLPV theories, we shall show the relations between the quantities 
in the two frames linked through the disformal transformation. 
Under the disformal transformation of the action (\ref{oriac}) with 
the Lagrangian (\ref{LH}), we obtain the action 
(\ref{Stra}) with $\hat{L}$ given by Eq.~(\ref{LH2}).

\subsection{Background quantities}

Let us first discuss the transformation of the background equations 
of motion (\ref{back1})-(\ref{back2}). 
Since $\hat{a}=\Omega a$, the Hubble parameter 
in the new frame, defined by 
$\hat{H}=\dot{\hat a}/(\hat{\bar N}{\hat a})$,
is related to $H=\dot{a}/(\bar{N}a)$ as
\be
\hat{H}=\frac{1}{\bar{\alpha}} \left( H+\frac{\omega}{\bar{N}} \right)\,,
\label{hatH}
\ee
where 
\be
\bar{\alpha} \equiv \frac{\hat{\bar N}}{\bar N}\,.
\label{alphadef}
\ee

In GLPV theories the function $\hat{{\cal F}}$ in the transformed 
frame is given by $\hat{{\cal F}}=\hat{A}_3+4\hat{H} \hat{A}_4
+6\hat{H}^2 \hat{A}_5$. 
Using the transformation laws (\ref{hatA3}), (\ref{hatA4}), (\ref{hatA5}) 
and (\ref{hatH}), it follows that 
\be
\hat{\cal F}=\frac{1}{\Omega^3}{\cal F}\,,
\label{calF}
\ee
where ${\cal F}=A_3+4HA_4+6H^2 A_5$.
Similarly, the transformation of the background Lagrangian 
$\bar{L}=A_2+3HA_3+6H^2A_4+6H^3 A_5$ is 
\be
\hat{\bar L}=\frac{1}{\Omega^3 \bar{\alpha}} \bar{L}\,.
\label{Lre}
\ee

Taking the $\hat{N}$ derivative of the coefficients (\ref{hatA2})-(\ref{hatA4}) 
and (\ref{hatA5}), we find that the background value of $\hat{L}_{,\hat{N}}$
is given by 
\be
\hat{L}_{,\hat{N}}=\frac{\bar{\beta}}{\Omega^3 \bar{\alpha}} 
\left[ L_{,N}-\bar{\mu} (\bar{L}-3H{\cal F})
+\frac{3\omega {\cal F}}{\bar{N}^2} (1+\bar{\mu} \bar{N}) \right]\,,
\label{LN}
\ee
where, in the square bracket of Eq.~(\ref{LN}), $N$ is replaced 
by $\bar{N}$ after taking the $N$ derivative, and 
\be
\bar{\beta} \equiv \frac{\partial N}{\partial \hat{N}}
\biggr|_{\hat{N}=\hat{\bar{N}}}\,,
\qquad
\bar{\mu} \equiv \frac{1}{\alpha} \frac{\partial \alpha}
{\partial N}\biggr|_{N=\bar{N}}\,.
\label{bemu}
\ee
{}From Eq.~(\ref{hatN}) there is the following relation 
\be
\bar{\beta}=\left(  \frac{\partial \hat{N}}{\partial N}\right)^{-1}\biggr|_{N=\bar{N}} 
=\frac{2\bar{\alpha}\bar{N}}{2\bar{N}\Omega^2-\Gamma_{,N}\dot{\phi}^2}\,,
\label{barbe}
\ee
so that $\bar{\mu}$, $\bar{\alpha}$, and $\bar{\beta}$ are 
related with each other, as
\be
\bar{\mu} \bar{N} = \frac{1}{\bar{\alpha} \bar{\beta}}-1\,.
\label{barmu}
\ee
On using Eqs.~(\ref{back1}) and (\ref{barmu}), 
Eq.~(\ref{LN}) reduces to 
\be
\hat{L}_{,\hat{N}}=\frac{1}{\Omega^3 \bar{\alpha}^2} 
\left( L_{,N}+\frac{3\omega {\cal F}}{\bar{N}^2} \right)\,.
\label{hatLN}
\ee
{}From Eqs.~(\ref{hatH}), (\ref{calF}), (\ref{Lre}), and 
(\ref{hatLN}) it follows that 
\ba
& &
\hat{\bar L}+\hat{\bar N} \hat{L}_{,\hat{N}}-3\hat{H}\hat{\cal F}
=\frac{1}{\Omega^3 \bar{\alpha}}
({\bar L}+{\bar N} L_{,N}-3H{\cal F})=0\,,
\label{backtra1} \\
& & 
\hat{\bar L}-\frac{\dot{\hat{\cal F}}}{\hat{\bar N}}
-3\hat{H}\hat{\cal F}=
\frac{1}{\Omega^3 \bar{\alpha}}
\left({\bar L}-\frac{\dot{\cal F}}{\bar N}-3H{\cal F} \right)=0\,,
\label{backtra2}
\ea
where we used the background Eqs.~(\ref{back1})-(\ref{back2}) 
in the original frame.
Equations (\ref{backtra1})-(\ref{backtra2}) are equivalent to those derived by 
varying the action $S=\int d^4 x \sqrt{-\hat{g}}\hat{L}$ 
in the transformed frame.

\subsection{Quantities associated with perturbations}

For scalar perturbations the second-order action 
in the transformed frame is given by 
$S^{(2)}=\int d^4 x\,\hat{{\cal L}}_2$, where 
\be
\hat{{\cal L}}_2=\hat{a}^3 \hat{q}_{\rs} 
\left[ \dot{\hat{\zeta}}^2-\frac{\hat{c}_{\rs}^2}{\hat{a}^2} 
(\partial \hat{\zeta})^2 \right]\,.
\label{L2tra}
\ee
Since the curvature perturbation $\zeta$ is invariant 
under the disformal transformation, the equivalence 
between the Lagrangian densities (\ref{L2}) and (\ref{L2tra})
implies the following relations
\ba
\hat{q}_{\rs} &=&
\frac{1}{\Omega^3}q_{\rs}\,,
\label{qstra} \\
\hat{c}_{\rs}^2 &=& \Omega^2 c_{\rs}^2\,.
\label{cstra} 
\ea
In what follows we shall derive these relations in GLPV theories 
by explicitly employing the transformation laws 
(\ref{hatA2})-(\ref{hatB5}) for the background variables 
appearing on the r.h.s. of Eqs.~(\ref{qs})-(\ref{cs}).

The quantities $L_{,\s}=-A_4-3HA_5$ and 
${\cal W}=A_{3,N}+4HA_{4,N}+6H^2A_{5,N}
-4H(A_4+3HA_5)/{\bar N}$
transform, respectively, as 
\be
\hat{L}_{,\hat{\s}}=\frac{\bar{\alpha}}{\Omega^3}L_{,\s}\,,
\qquad
\hat{\cal W}=\frac{\bar{\beta}}{\Omega^3}{\cal W}\,.
\label{LsW}
\ee
We recall that the transformation of $L_{,N}$ 
is given by Eq.~(\ref{hatLN}). 
For the computation of the quantity $\hat{L}_{,\hat{N}\hat{N}}$, 
we take the second derivatives of Eqs.~(\ref{hatA2})-(\ref{hatA4}) 
and (\ref{hatA5}) with respect to $\hat{N}$ and then substitute 
the relations $A_2=6H^2A_4+12H^3A_5-\bar{N}L_{,N}$ 
[which comes from Eq.~(\ref{back1})], 
$A_{2,N}=L_{,N}-3HA_{3,N}-6H^2 A_{4,N}-6H^3 A_{5,N}$,
$A_{2,NN}=L_{,NN}-3HA_{3,NN}-6H^2 A_{4,NN}-6H^3 A_{5,NN}$, 
$A_4=-L_{,\s}-3HA_5$, 
$A_{3,N}={\cal W}-4HA_{4,N}-6H^2A_{5,N}-4L_{,\s}H/{\bar N}$, 
and Eq.~(\ref{barmu}) into the expression of $\hat{L}_{,\hat{N}\hat{N}}$.
This process leads to 
\ba
\hat{L}_{,\hat{N}\hat{N}}
&=&\frac{1}{\Omega^3 \bar{\alpha}^3 \bar{N}^4}
\biggl[ \bar{\alpha} \bar{\beta} \bar{N}^3 \left\{ \bar{\alpha} \bar{\beta} 
\bar{N}L_{,NN}-6H(\bar{\alpha} \bar{\beta}-1){\cal W} \right\}
+2(\bar{\alpha}^2 \bar{\beta}^2-1) \bar{N}^2
(\bar{N}L_{,N}+6H^2L_{,\s}) \nonumber \\
& &~~~~~~~~~~~~
-6\omega \left( \bar{N}{\cal F}+4\bar{N}HL_{,\s}
-\bar{\alpha} \bar{\beta} \bar{N}^2{\cal W}
+2L_{,\s} \omega \right) \biggr]\,.
\label{LNN}
\ea
On using Eqs.~(\ref{hatH}), (\ref{hatLN}), (\ref{LsW}), and (\ref{LNN}), 
it follows that 
\ba
&& 
4\hat{L}_{,\hat{\s}} ( 2\hat{\bar{N}}\hat{L}_{,\hat{N}}
+\hat{\bar{N}}^2\hat{L}_{,\hat{N}\hat{N}} )
+3( \hat{\bar{N}}\hat{\cal W}-4\hat{H} \hat{L}_{,\hat{\s}} )^2 \nonumber \\
& &=\frac{\bar{\alpha}^2 \bar{\beta}^2}{\Omega^6} 
[4L_{,\s}(2\bar{N}L_{,N}+\bar{N}^2L_{,NN})+
3(\bar{N}{\cal W}-4HL_{,\s})^2]\,.
\label{Lsre}
\ea
{}From Eqs.~(\ref{LsW}) and (\ref{Lsre}) the quantity 
$q_{\rs}$ defined by Eq.~(\ref{qs}) indeed obeys 
the transformation law (\ref{qstra}).

In GLPV theories the quantities ${\cal E}$ and ${\cal M}$ are given, 
respectively, by ${\cal E}=B_4+\dot{B}_5/(2\bar{N})$ and 
${\cal M}=4L_{,\s} (B_4+\bar{N}B_{4,N}-H \bar{N} B_{5,N}/2)
/(\bar{N}{\cal W})$. Employing Eqs.~(\ref{hatB4}), (\ref{hatB5}), 
(\ref{hatH}), (\ref{barmu}), and (\ref{LsW}), 
we find that ${\cal E}$ and ${\cal M}$ transform as 
\ba
\hat{\cal E} &=& \frac{1}{\Omega \bar{\alpha}} {\cal E}\,,\label{E}\\
\hat{\cal M} &=& \frac{1}{\Omega}{\cal M}\,.
\label{M}
\ea
{}From the definition (\ref{cs}) of $c_{\rs}^2$ together 
with Eqs.~(\ref{hatH}), (\ref{qstra}), (\ref{E}), and (\ref{M}), 
it follows that the scalar propagation speed squared 
transforms as Eq.~(\ref{cstra}).

The quantities $q_{\rt}$ and $c_{\rt}^2$ appearing in the 
tensor perturbation equation of motion are given, 
respectively, by Eqs.~(\ref{qtdef}) and (\ref{ctdef}). 
{}From Eqs.~(\ref{LsW}) and (\ref{E}) 
they transform as
\ba
\hat{q}_{\rt} &=&
\frac{1}{\Omega^3}q_{\rt}\,,
\label{qttra} \\
\hat{c}_{\rt}^2 &=& \Omega^2 c_{\rt}^2\,.
\label{cttra} 
\ea
These properties also follow from the equivalence of the 
second-order tensor Lagrangian densities ${\cal L}_2^{(h)}$ 
and $\hat{{\cal L}}_2^{(h)}$ in the two frames 
together with the invariance (\ref{gamma1}) of tensor perturbations.

\section{Next-to-leading order inflationary power spectra} \label{powerspssec}

We derive the power spectra of scalar and tensor perturbations 
generated during inflation up to next-to-leading order in slow-roll 
and show that they are invariant under the disformal transformation.

We shall consider the quasi de-Sitter background on which 
the expansion rate defined by 
\be
h \equiv \bar{N}H=\frac{\dot{a}}{a}
\ee
is nearly constant. In this case the slow-roll parameter
\be
\epsilon_h \equiv -\frac{\dot{h}}{h^2}\,
\ee
is much smaller than 1. 
The conformal time $\tau \equiv \int a^{-1}dt$ 
is approximately given by $\tau \simeq -1/(ah)$ on the quasi de-Sitter 
background (where we set the integration constant 0), 
so the asymptotic past and future correspond 
to $\tau \to -\infty$ and $\tau \to 0$, 
respectively. If there is a function $f(\tau)$ that varies 
slowly during inflation, the expansion of $f(\tau)$ 
around a chosen time $\tau_k$ is \cite{Chen}
\be
f(\tau) \simeq f(\tau_k) \left[ 
1-\delta_f (\tau_k) \ln \frac{\tau}{\tau_k} \right]\,, 
\qquad
\delta_f \equiv \frac{\dot{f}}{hf}\,,
\label{frun}
\ee
where $\delta_f$ is regarded as a slow-roll parameter which 
is the same order as $\epsilon_h$.

\subsection{The scalar power spectrum}

The Lagrangian density for scalar perturbations is given 
by Eq.~(\ref{L2}).
The curvature perturbation $\zeta({\bm x}, \tau)$ in real space 
can be expressed in terms of the Fourier components 
$\zeta (k,\tau)$ with the comoving wave number ${\bm k}$, as
\begin{equation}
\zeta({\bm x}, \tau)=\int \frac{d^3 k}{(2\pi)^{3/2}}
e^{i {\bm k}\cdot {\bm x}} \tilde{\zeta}({\bm k},\tau)\,,\qquad
\tilde{\zeta}({\bm k},\tau)=\zeta(k,\tau)a({\bm k})
+\zeta^{*}(k,\tau) a^\dagger(-{\bm k})\,,
\label{zeta0def}
\end{equation}
where $a({\bm k})$ and $a^\dagger({\bm k'})$ are 
the annihilation and creation operators satisfying 
the commutation relation 
$[a({\bm k}), a^\dagger({\bm k'}) ]=\delta^{(3)} ({\bm k}-{\bm k}')$.
A rescaled field defined by 
\be
v_{\rs} (k,\tau)=z_{\rs}\,\zeta(k,\tau)\,,\qquad
z_{\rs}=a\sqrt{2q_{\rs}}
\ee
corresponds to a canonical scalar field associated with 
the quantization procedure.
{}From Eq.~(\ref{scaeq}) it follows that 
\be
v_{\rs}''+\left( c_{\rs}^2 k^2 -\frac{z_{\rs}''}{z_{\rs}} \right)v_{\rs}=0\,,
\label{vseq}
\ee
where a prime represents the derivative with respect to $\tau$.

We introduce the slow-roll parameters
\be
\epsilon_{\rs} \equiv \frac{\dot{q}_{\rs}}{hq_{\rs}}\,,
\qquad s_{\rs} \equiv \frac{\dot{c}_{\rs}}{hc_{\rs}}\,,
\ee
and pick up the terms up to next-to-leading order 
in slow-roll under the assumption that 
$|\epsilon_{\rs}|$ and $|s_{\rs}|$
are much smaller than 1 during inflation.
The quantity $z_{\rs}''/z_{\rs}$ in Eq.~(\ref{vseq}) can be 
estimated as
\be
\frac{z_{\rs}''}{z_{\rs}} \simeq 2(ah)^2 \left( 1-\frac12 \epsilon_h
+\frac34 \epsilon_{\rs} \right)\,.
\ee

The sound-horizon crossing corresponds to the epoch 
characterized by $c_{\rs}k=ah$.
Defining the dimensionless variable
\be
y_{\rs} \equiv \frac{c_{\rs}k}{ah}\,,
\ee
Eq.~(\ref{vseq}) can be written as
\be
(1-2\epsilon_h-2s_{\rs})y_{\rs}^2 \frac{d^2v_{\rs}}
{dy_{\rs}^2}-s_{\rs}y_{\rs} \frac{dv_{\rs}}{dy_{\rs}}
+\left( y_{\rs}^2-2+\epsilon_h-\frac32 \epsilon_{\rs}
\right)v_{\rs}=0\,.
\ee
The solution to this equation is given by 
\be
v_{\rs}
= y_{\rs}^{(1+s_{\rs})/2} \left\{ \alpha_k 
H_{\nu}^{(1)}[(1+\epsilon_h+s_{\rs})y_{\rs}] 
+\beta_k H_{\nu}^{(2)}[(1+\epsilon_h+s_{\rs}) 
y_{\rs}] \right\}\,,
\label{vso} 
\ee
where $\alpha_k$ and $\beta_k$ are integration constants,
$H_{\nu}^{(1)}(x)$ and $H_{\nu}^{(2)}(x)$ are 
Hankel functions of the first and second kinds 
respectively, and
\be
\nu=\frac32+\epsilon_{h}+\frac12 \epsilon_{\rs}
+\frac32 s_{\rs}\,.
\ee
For the derivation of the solution (\ref{vso}), we have ignored 
the slow-roll parameters higher than second order. 

The quantities $h$, $q_{\rs}$, and $c_{\rs}$ have runnings according 
to Eq.~(\ref{frun}), i.e., $\delta_f=-\epsilon_h, \epsilon_{\rs}, s_{\rm s}$ 
for $f=h,q_{\rs},c_{\rs}$, respectively.
Provided that $|\delta_f| \ll 1$, the function $1-\delta_f \ln (\tau/\tau_k)$ 
is approximately equivalent to $y_s^{-\delta_f}$.
Hence the runnings of $h$, $q_{\rs}$, and $c_{\rs}$ 
can be quantified as
\be
h=h_k\,y_{\rs}^{\epsilon_h}\,,\qquad
q_{\rs}=q_{\rs k}\,y_{\rs}^{-\epsilon_{\rs}}\,,\qquad
c_{\rs}=c_{\rs k}\,y_{\rs}^{-s_{\rs}}\,,
\label{hqc}
\ee
where the lower indices $k$ represent the values at $y_\rs=1$.

The positive-frequency solution satisfying the Wronskian condition
$v_{\rs}v_{\rs}^{*'}-v_{\rs}^{*}v_{\rs}^{'}=i$ in the asymptotic 
past ($y_{\rs} \to \infty$) corresponds to the coefficients
\be
\alpha_k =-\frac12 \sqrt{\frac{\pi}{c_{\rs k}\,k}}\,
\left(1+\frac12 \epsilon_{h}+\frac12 s_{\rs} \right)\,,
\qquad \beta_k=0\,,
\label{coeff}
\ee
where we exploited the last relation of Eq.~(\ref{hqc}).
Substituting Eq.~(\ref{coeff}) into Eq.~(\ref{vso}) and 
using the first and second relations of Eq.~(\ref{hqc}), 
the solution to the curvature perturbation $\zeta=v_{\rs}/z_{\rs}$
long after the sound horizon crossing ($y_{\rs} \to 0$)
is given by 
\be
\zeta(k,0)=i \frac{2^{\nu} \Gamma (\nu)}{\sqrt{8\pi q_{\rs k}}}
\frac{1-\epsilon_h-s_{\rs}}{(c_{\rs k}k)^{3/2}}h_k\,,
\label{zeta0}
\ee
where $\Gamma (\nu)$ is the Gamma function. 

The power spectrum ${\cal P}_{\zeta}$ is defined by 
the vacuum expectation value of the two-point correlation 
function of $\zeta$, as
\begin{equation}
\langle 0| \tilde{\zeta}({\bm k}_1,0) \tilde{\zeta}({\bm k}_2,0)|0 \rangle
=\frac{2\pi^2}{k_1^3} \delta^{(3)}({\bm k}_1+{\bm k}_2) 
{\cal P}_{\zeta}(k_1)\,.
\end{equation}
Employing the solution (\ref{zeta0}) and expanding the Gamma function 
around $\nu=3/2$, we obtain the next-to-leading order scalar power spectrum
\be
{\cal P}_{\zeta}(k)
=\frac{h^2}{8\pi^2 q_{\rs}c_{\rs}^3} 
\left[ 1-2(C+1) \epsilon_h-C\epsilon_{\rs}
-(3C+2)s_{\rs} \right] \biggr|_{c_{\rs}k=ah}\,,
\label{Pzeta}
\ee
where $C=\gamma-2+\ln 2=-0.729637...$ 
($\gamma$ is the Euler-Mascheroni constant).

\subsection{The tensor power spectrum}

The next-to-leading order power spectrum of tensor perturbations
was recently derived in Ref.~\cite{DeTsuji14}.
Here, we briefly summarize the result by explicitly taking into 
account the lapse function $\bar{N}$. We first 
expand $\gamma_{ij} ({\bm x}, \tau)$ into the Fourier series as
\be
\gamma_{ij} ({\bm x}, \tau)=\int \frac{d^3k}{(2\pi)^{3/2}}
e^{i {\bm k} \cdot {\bm x}}\tilde{\gamma}_{ij}({\bm k}, \tau)\,,\qquad
\tilde{\gamma}_{ij}({\bm k}, \tau)= \sum_{\lambda=+,\times} 
\tilde{h}_{\lambda} ({\bm k}, \tau) e_{ij}^{(\lambda)} ({\bm k})\,,
\label{gamex1}
\ee
where $e^{(\lambda)}_{ij} (\bm{k})$ are the transverse and 
traceless polarization tensors satisfying the normalization
$e^{(\lambda)}_{ij} (\bm{k})  e^{*(\lambda')}_{ij} (\bm{k}) 
= \delta_{\lambda \lambda'}$.
The annihilation and creation operators
$a_{\lambda} ({\bm k})$ and $a_{\lambda}^{\dagger} 
({\bm k}')$ obey the commutation relation 
$[a_{\lambda} ({\bm k}),a_{\lambda'}^{\dagger} 
({\bm k}')]=\delta_{\lambda \lambda'}\delta^{(3)}
({\bm k}-{\bm k}')$. 
The Fourier mode $\tilde{h}_{\lambda} ({\bm k}, \tau)$ 
is expressed in the form
\be
\tilde{h}_{\lambda} ({\bm k}, \tau)=
h_{\lambda} (k, \tau) a_{\lambda} ({\bm k})+
h_{\lambda}^* (k, \tau) a_{\lambda}^{\dagger} (-{\bm k})\,.
\label{gamex2}
\ee

Defining a canonically normalized field $v_{\lambda} (k, \tau)$ as 
\be
v_{\lambda} (k, \tau)\equiv z_{\rt} h_{\lambda}(k, \tau)\,,\qquad
z_{\rt} \equiv a \sqrt{2q_{\rt}}\,,
\label{vz}
\ee
Eq.~(\ref{tensoreq}) reduces to
\be
v_{\lambda}''+\left( c_{\rt}^2 k^2 -\frac{z_{\rt}''}{z_{\rt}} \right)
v_{\lambda}=0\,.
\label{vteq}
\ee
We can derive the solution to this equation by introducing a 
dimensionless parameter $y_{\rt} \equiv c_{\rt} k/(ah)$. 
Following the similar procedure to that performed for scalar perturbations, 
the solution to $h_{\lambda}$ in the regime $\tau \to 0$ is given by 
\be
h_{\lambda}(k,0)=i \frac{2^{\nu_{\rt}} \Gamma (\nu_\rt)}{\sqrt{8\pi q_{\rt k}}}
\frac{1-\epsilon_h-s_{\rt}}{(c_{\rt k}k)^{3/2}}h_{k}\,,
\label{h0}
\ee
where the lower index $k$ represents the values 
at $y_{\rt}=1$, and 
\be
\nu_{\rt}=\frac32+\epsilon_{h}+\frac12\epsilon_{\rt}
+\frac32 s_{\rt}\,,\qquad
\epsilon_{\rt} \equiv \frac{\dot{q}_{\rt}}{hq_{\rt}}\,,
\qquad s_{\rt} \equiv \frac{\dot{c}_{\rt}}{hc_{\rt}}\,.
\ee
Defining the tensor power spectrum 
${\cal P}_h$ as 
\be
\langle 0| \tilde{\gamma}_{ij} ({\bm k}_1, 0) 
 \tilde{\gamma}_{ij} ({\bm k}_2, 0)  |0
\rangle=\frac{2\pi^2}{k_1^3}\delta^{(3)} 
({\bm k}_1+{\bm k}_2) {\cal P}_h (k_1)\,,
\ee
it follows that 
\be
{\cal P}_h(k)=\frac{h^2}{4\pi^2 q_{\rt}c_{\rt}^3} 
\left[ 1-2(C+1) \epsilon_h-C\epsilon_{\rt}
-(3C+2)s_{\rt} \right] \biggr|_{c_{\rt}k=ah}\,.
\label{Ph}
\ee
We note that the tensor power spectrum should be 
evaluated at $c_{\rt}k=ah$, which is generally different 
from the moment $c_{\rs}k=ah$ for
the sound horizon crossing of scalar perturbations.

\subsection{Spectral indices and the tensor-to-scalar ratio}

The spectral index of scalar perturbations is defined by
\be
n_{\rs}-1=\frac{d\ln {\cal P}_{\zeta}(k)}{d \ln k}\biggr|_{c_{\rs}k=ah}\,,
\ee
which reduces to $n_{\rs}-1=(\dot{{\cal P}}_\zeta/{\cal P}_{\zeta})(dt/d\ln k)|_{c_{\rs}k=ah}$.
Taking the time derivative of Eq.~(\ref{Pzeta}) and using 
the relation $d\ln k/dt|_{c_{\rs}k=ah}=h_k(1-\epsilon_h-s_{\rs})$, 
we obtain
\ba
n_{\rs}-1 
&=& -2\epsilon_h-\epsilon_{\rs}-3s_{\rs}
-2\epsilon_h^2-5\epsilon_hs_{\rs}-\epsilon_h \epsilon_{\rs}
-\epsilon_{\rs}s_{\rs}-3s_{\rs}^2 \nonumber \\
& & -2(C+1)\epsilon_h \eta_h-C \epsilon_{\rs} \eta_{\rs}
-(3C+2)s_{\rs} \mu_{\rs}\,|_{c_{\rs}k=ah}\,,
\label{ns}
\ea
where 
\be
\eta_h \equiv \frac{\dot{\epsilon}_h}{h \epsilon_{h}}\,,
\qquad
\eta_{\rs} \equiv \frac{\dot{\epsilon}_{\rs}}{h \epsilon_{\rs}}\,,
\qquad
\mu_{\rs} \equiv \frac{\dot{s}_{\rs}}{hs_{\rs}}\,.
\ee

We also introduce the tensor spectral index 
\be
n_{\rt}=\frac{d\ln {\cal P}_{h}(k)}{d \ln k}\biggr|_{c_{\rt}k=ah}\,.
\ee
{}From Eq.~(\ref{Ph}) it follows that 
\ba
n_{\rt}
&=& -2\epsilon_h-\epsilon_{\rt}-3s_{\rt}
-2\epsilon_h^2-5\epsilon_h s_{\rt}-\epsilon_h \epsilon_{\rt}
-\epsilon_{\rt} s_{\rt}-3s_{\rt}^2 \nonumber \\
& & -2(C+1)\epsilon_h \eta_h-C \epsilon_{\rt} \eta_{\rt}
-(3C+2)s_{\rt} \mu_{\rt}\,|_{c_{\rt}k=ah}\,,
\label{nt}
\ea
where 
\be
\eta_{\rt} \equiv \frac{\dot{\epsilon}_{\rt}}{h \epsilon_{\rt}}\,,
\qquad
\mu_{\rt} \equiv \frac{\dot{s}_{\rt}}{hs_{\rt}}\,.
\ee

The scalar and tensor spectra (\ref{Pzeta}) and (\ref{Ph}) 
are computed at $c_{\rs}k=ah$ and $c_{\rt}k=ah$, respectively.
The moment for the evaluation of the tensor-to-scalar ratio 
$r={\cal P}_{h}(k)/{\cal P}_{\zeta}(k)$ is 
different depending on the values of $c_{\rs}$ and $c_{\rt}$.

If $c_{\rs}<c_{\rt}$, the sound horizon crossing 
for scalar perturbations occurs at $y_{\rs}=c_{\rs}k/(ah)=1$ 
(denoted as $\tau=\tau_{\rs}$), 
whereas for tensor perturbations it corresponds to 
the later time characterized by $y_{\rs}=c_{\rs}/c_{\rt}<1$ 
(denoted as $\tau=\tau_{\rt}$).
Since the tensor perturbation evolves during the epoch 
$c_{\rs}/c_{\rt}<y_{\rs}<1$ and it starts to be frozen 
in the regime  $y_{\rs}<c_{\rs}/c_{\rt}$, the tensor-to-scalar ratio 
should be evaluated at $y_{\rs}=c_{\rs}/c_{\rt}$, i.e., 
$c_{\rt}k=ah$. 

{}From Eq.~(\ref{frun}), any time-dependent function $f(\tau)$ can be 
expanded as $f(\tau) \simeq f(\tau_k)x^{-\delta_f}$ 
for $|\delta_f| \ll 1$, where $x \equiv k/(ah)$ and $\tau_k$ is the 
instant at $x=1$.
Since $x=1/c_{\rs}$ at $\tau=\tau_{\rs}$ and $x=1/c_{\rt}$ 
at $\tau=\tau_{\rt}$, it follows that 
$f(\tau_{\rs}) \simeq f(\tau_k)c_{\rs}^{\delta_f}$ and 
$f(\tau_{\rt}) \simeq f(\tau_k)c_{\rt}^{\delta_f}$. 
In this case we have
$f(\tau_{\rs}) \simeq \left[ 1+\delta_f \ln (c_{\rs}/c_{\rt})
\right] f(\tau_{\rt})$, so that  
$h(\tau_{\rs})=\left[ 1-\epsilon_h \ln (c_{\rs}/c_{\rt})
\right] h (\tau_{\rt})$,
$q_{\rs}(\tau_{\rs})=\left[ 1+\epsilon_{\rs} \ln (c_{\rs}/c_{\rt}) 
\right] q_{\rs} (\tau_{\rt})$, and 
$c_{\rs}(\tau_{\rs})=\left[ 1+s_{\rs} \ln (c_{\rs}/c_{\rt})  
\right] c_{\rs} (\tau_{\rt})$.
Substituting these relations into Eq.~(\ref{Pzeta}), the next-to-leading 
order scalar power spectrum can be written in the form 
\be
{\cal P}_{\zeta}(k)
=\frac{h^2}{8\pi^2 q_{\rs}c_{\rs}^3} 
\left[ 1-2(C+1) \epsilon_h-C\epsilon_{\rs}
-(3C+2)s_{\rs} 
-(2\epsilon_h+\epsilon_{\rs}+3s_{\rs}) \ln 
\frac{c_{\rs}}{c_{\rt}}
\right] \biggr|_{c_{\rt}k=ah}\,.
\label{Pzeta2}
\ee
On using Eq.~(\ref{Ph}), the tensor-to-scalar ratio evaluated at 
$c_{\rt}k=ah$ reads
\be
r=2\frac{q_{\rs}c_{\rs}^3}{q_{\rt}c_{\rt}^3} 
\left[ 1-C(\epsilon_{\rt}-\epsilon_{\rs})
-(3C+2)(s_{\rt}-s_{\rs})
+(2\epsilon_h+\epsilon_{\rs}+3s_{\rs}) \ln 
\frac{c_{\rs}}{c_{\rt}}
\right] \biggr|_{c_{\rt}k=ah}\,,
\label{ratio}
\ee
which is valid for $c_{\rs}<c_{\rt}$.

If $c_{\rs}>c_{\rt}$, then the scalar perturbation is frozen at a later epoch 
relative to the tensor perturbation. 
Following the similar procedure to that given above, the tensor 
power spectrum and the tensor-to-scalar ratio at $c_{\rs}k=ah$ 
are given, respectively, by 
\be
{\cal P}_{h}(k)
=\frac{h^2}{4\pi^2 q_{\rt}c_{\rt}^3} 
\left[ 1-2(C+1) \epsilon_h-C\epsilon_{\rt}
-(3C+2)s_{\rt} 
+(2\epsilon_h+\epsilon_{\rt}+3s_{\rt}) \ln 
\frac{c_{\rs}}{c_{\rt}}
\right] \biggr|_{c_{\rs}k=ah}\,.
\label{Pzeta3}
\ee
and 
\be
r=2\frac{q_{\rs}c_{\rs}^3}{q_{\rt}c_{\rt}^3} 
\left[ 1-C(\epsilon_{\rt}-\epsilon_{\rs})
-(3C+2)(s_{\rt}-s_{\rs})
+(2\epsilon_h+\epsilon_{\rt}+3s_{\rt}) \ln 
\frac{c_{\rs}}{c_{\rt}}
\right] \biggr|_{c_{\rs}k=ah}\,.
\label{ratio2}
\ee
Compared to Eq.~(\ref{ratio}), the difference arises for the 
two terms in front of the factor $\ln (c_\rs/c_\rt)$.

\subsection{Equivalence under the disformal transformation}

In what follows we show the equivalence of inflationary observables 
under the disformal transformation by using the relations between 
the quantities appearing in the scalar and tensor power spectra.
Defining the expansion rate $\hat{h}=\hat{\bar{N}}\hat{H}$ in the 
transformed frame, we obtain the following relation from 
Eq.~(\ref{hatH}): 
\be
\hat{h}=h(1+\epsilon_{\Omega})\,,\qquad
\epsilon_{\Omega} \equiv \frac{\omega}{h}
=\frac{\dot{\Omega}}{h \Omega}\,.
\label{hath}
\ee
Then, the slow-roll parameter $\hat{\epsilon}_h=-\dot{\hat{h}}/{\hat{h}^2}$ 
can be expressed as 
\be
\hat{\epsilon}_h=\epsilon_h-\epsilon_h \epsilon_{\Omega}
-\frac{\dot{\epsilon}_{\Omega}}{h}+O(\epsilon_h^3)\,.
\label{hateph}
\ee
On using Eqs.~(\ref{qstra}), (\ref{cstra}), (\ref{qttra}) and (\ref{cttra}), 
the quantities $\hat{\epsilon}_{\rs}=\dot{\hat{q}}_{\rs}/(\hat{h}\hat{q}_{\rs})$, 
$\hat{s}_{\rs}=\dot{\hat{{c}}}_{\rs}/(\hat{h}\hat{c}_{\rs})$, 
$\hat{\epsilon}_{\rt}=\dot{\hat{q}}_{\rt}/(\hat{h}\hat{q}_{\rt})$, 
$\hat{s}_{\rt}=\dot{\hat{{c}}}_{\rt}/(\hat{h}\hat{c}_{\rt})$ are 
given, respectively, by 
\ba
& &\hat{\epsilon}_{\rs}=\epsilon_{\rs}-3\epsilon_{\Omega}
-\epsilon_{\rs} \epsilon_{\Omega}+3\epsilon_{\Omega}^2
+O(\epsilon_h^3)\,,\qquad
\hat{s}_{\rs}=s_{\rs}+\epsilon_{\Omega}-s_{\rs} \epsilon_{\Omega}
-\epsilon_{\Omega}^2+O(\epsilon_h^3)\,,\label{hateprs} \\
& &\hat{\epsilon}_{\rt}=\epsilon_{\rt}-3\epsilon_{\Omega}
-\epsilon_{\rt} \epsilon_{\Omega}+3\epsilon_{\Omega}^2
+O(\epsilon_h^3)\,,\qquad
\hat{s}_{\rt}=s_{\rt}+\epsilon_{\Omega}-s_{\rt} \epsilon_{\Omega}
-\epsilon_{\Omega}^2+O(\epsilon_h^3)\,.
\label{hateprt}
\ea

The scalar power spectrum in the transformed frame is given by 
\be
\hat{{\cal P}}_{\zeta}(k)
=\frac{\hat{h}^2}{8\pi^2 \hat{q}_{\rs}\hat{c}_{\rs}^3} 
\left[ 1-2(C+1) \hat{\epsilon}_h-C\hat{\epsilon}_{\rs}
-(3C+2)\hat{s}_{\rs} \right] |_{\hat{c}_{\rs}k=\hat{a}\hat{h}} \,.
\label{Pzetatra}
\ee
Substituting Eqs.~(\ref{qstra}), (\ref{cstra}), (\ref{hath}), (\ref{hateph}), 
and (\ref{hateprs}) into Eq.~(\ref{Pzetatra}), it follows that 
\be
\hat{{\cal P}}_{\zeta}(k)={\cal P}_{\zeta}(k)\,,
\label{zetaeq}
\ee
up to next-to-leading order in slow-roll.
{}From Eqs.~(\ref{hata}), (\ref{cstra}), and (\ref{hath}) 
the moment $\hat{c}_{\rs}k=\hat{a}\hat{h}$ corresponds to 
$c_{\rs}k=ah(1+\epsilon_{\Omega})$, which differs from 
$c_{\rs}k=ah$ due to the presence of the factor $\epsilon_{\Omega}$.
However, this difference does not affect the next-to-leading order 
power spectrum because the variations of the quantities like 
$h, q_{\rs}, c_{\rs}$ are quantified according to Eq.~(\ref{hqc}).
Since the variable $\hat{y}_{\rs}=\hat{c}_{\rs}k/(\hat{a}\hat{h})$ 
is related to $y_{\rs}=c_{\rs}k/(ah)$ as $\hat{y}_{\rs}=y_{\rs}/(1+\epsilon_{\Omega})$, 
the quantities like $y_{\rs}^{\epsilon_{h}}$ only give rise to second-order 
slow-roll corrections. 

{}From Eqs.~(\ref{qttra}), (\ref{cttra}), (\ref{hath}), (\ref{hateph}), and 
(\ref{hateprt}), it also follows that 
\be
\hat{{\cal P}}_{h}(k)={\cal P}_{h}(k)\,,
\label{heq}
\ee
up to next-to-leading order.

The spectral index of scalar perturbations in the transformed frame 
is given by 
\ba
\hat{n}_{\rs}-1 
&=& -2\hat{\epsilon}_h-\hat{\epsilon}_{\rs}-3\hat{s}_{\rs}
-2\hat{\epsilon}_h^2-5\hat{\epsilon}_h \hat{s}_{\rs}-\hat{\epsilon}_h 
\hat{\epsilon}_{\rs}
-\hat{\epsilon}_{\rs} \hat{s}_{\rs}-3\hat{s}_{\rs}^2 \nonumber \\
& & -2(C+1)\hat{\epsilon}_h \hat{\eta}_h-C \hat{\epsilon}_{\rs} \hat{\eta}_{\rs}
-(3C+2)\hat{s}_{\rs} \hat{\mu}_{\rs}|_{\hat{c}_{\rs}k=\hat{a}\hat{h}}\,.
\label{hatns}
\ea
We substitute Eqs.~(\ref{hateph}) and (\ref{hateprs}) into Eq.~(\ref{hatns}) and 
use the properties that the last two terms of 
Eq.~(\ref{hatns}) are given by 
$-C\hat{\epsilon}_{\rs}\hat{\eta}_{\rs}
=-C\epsilon_{\rs}\eta_{\rs}+3C\dot{\epsilon}_{\Omega}/h+O(\epsilon_h^3)$ and 
$-(3C+2)\hat{s}_{\rs}\hat{\mu}_{\rs}
=-(3C+2)s_{\rs}\mu_{\rs}-(3C+2) \dot{\epsilon}_{\Omega}/h+O(\epsilon_h^3)$, 
respectively. Then the terms involving $\epsilon_{\Omega}$ 
and $\dot{\epsilon}_{\Omega}$ vanish, so that we finally obtain 
\ba
\hat{n}_{\rs}=n_{\rs}.
\ea
Similarly, we can show the equivalence of the tensor spectral index:
\ba
\hat{n}_{\rt}=n_{\rt}.
\ea
Substituting Eqs.~(\ref{qstra}), (\ref{cstra}), (\ref{qttra}), (\ref{cttra}), 
(\ref{hateph}),  (\ref{hateprs}) and (\ref{hateprt}) into 
Eqs.~(\ref{ratio}) and (\ref{ratio2}), 
it also follows that 
\ba
\hat{r}=r\,,
\ea
up to next-to-leading order in slow roll.

\section{Einstein frame} \label{einsteinsec}

The action (\ref{LH}) of GLPV theories can be transformed to that 
in the so-called Einstein frame under the disformal transformation. 
As we will see below, the existence of the Einstein frame is related to a 
General Relativistic form of the inflationary tensor 
power spectrum.

\subsection{Inflationary power spectra in the Einstein frame}

{}From Eq.~(\ref{Ph}) the tensor power spectrum in the transformed 
frame is given by 
\be
\hat{{\cal P}}_h (k)=\frac{\hat{\bar N}^2 \hat{H}^2}
{4\pi^2 \hat{q}_{\rt k} \hat{c}_{\rt k}^3} \left[ 
1-2(C+1) \hat{\epsilon}_h-C\hat{\epsilon}_{\rt}
-(3C+2)\hat{s}_{\rt} \right]|_{\hat{c}_{\rt}k=\hat{a}\hat{h}}\,,
\label{hPh}
\ee
where $\hat{H}=\dot{\hat{a}}/(\hat{\bar N} \hat{a})$ is the 
Hubble parameter in the new frame. 
Note that the lapse $\hat{\bar N}$ has the same dimension as the 
tensor propagation speed $\hat{c}_{\rt k}$. Let us consider the case 
in which the quantities $\hat{c}_{\rt k}$ and $\hat{q}_{\rt k}$, 
after the disformal transformation, are given by 
\ba
\hat{c}_{\rt k} &=& \hat{\bar N}\,,\label{ck}\\
\hat{q}_{\rt k} &=& \frac{M_{\rm pl}^2}{8\hat{\bar N}}\,,
\label{qk}
\ea
where $M_{\rm pl}$ is the reduced Planck mass.
On using Eqs.~(\ref{hatN}), (\ref{qttra}), and (\ref{cttra}), 
the choices (\ref{ck}) and (\ref{qk}) correspond to
\be
\Omega^2=\frac{8q_{\rt k} c_{\rt k}}{M_{\rm pl}^2}\,,\qquad
\Gamma=\frac{8q_{\rt k} c_{\rt k}}{M_{\rm pl}^2}
\frac{c_{\rt k}^2-\bar{N}^2}{\bar{N^2}X}\,,
\label{OmeGam}
\ee
where $c_{\rt k}$ and $q_{\rt k}$ should be 
evaluated on the background such that the kinetic 
term $X$ appearing in these quantities is replaced 
by $\bar{X}=-\bar{N}^{-2}\dot{\phi}^2$.
For the specific case in which $\Omega^2$ is equivalent to 1, 
the factor $\Gamma$ in Eq.~(\ref{OmeGam}) 
matches with the results derived in Refs.~\cite{Cre,DeTsuji14} 
by setting $\bar{N}=1$.

For the choice (\ref{OmeGam}) the slow-roll parameters 
$\hat{\epsilon}_{\rt}$ and $\hat{s}_{\rt}$ in the transformed frame 
have the following relations
\be
\hat{\epsilon}_{\rt}=-\hat{s}_{\rt}\,.
\label{eprt}
\ee
Since $\hat{h}=\hat{\bar N}\hat{H}$, the slow-roll parameter 
$\hat{\epsilon}_h$ reads
\be
\hat{\epsilon}_h=\hat{\epsilon}_{H}-\hat{s}_{\rt}\,,
\label{hateph2}
\ee
where 
\be
\hat{\epsilon}_H \equiv -\frac{1}{\hat{\bar N}}
\frac{\dot{\hat H}}{\hat{H}^2}\,.
\label{epH}
\ee
Substituting Eqs.~(\ref{ck}), (\ref{qk}), (\ref{eprt}), and (\ref{hateph2}) 
into Eq.~(\ref{hPh}), it follows that 
\be
\hat{\cal P}_h (k)=\frac{2\hat{H}^2}{\pi^2 M_{\rm pl}^2} 
\left[ 1-2(C+1) \hat{\epsilon}_H \right]|_{k=\hat{a}\hat{H}}\,,
\label{Phein}
\ee
where we used the fact that the condition $\hat{c}_{\rt}k=\hat{a}\hat{h}$ 
translates to $k=\hat{a} \hat{H}$ by using Eq.~(\ref{ck}) with 
$\hat{h}=\hat{\bar{N}}\hat{H}$.
The spectrum (\ref{Phein}) is equivalent to the next-to-leading order 
tensor power spectrum in GR \cite{Stewart}. 
As we will see in Sec.~\ref{Einred}, the metric frame derived under the
disformal transformation with the factors (\ref{OmeGam}) 
can be regarded as the Einstein frame 
in which the function $\hat{A}_4$ involves the term 
$-M_{\rm pl}^2/2$, i.e., the Einstein-Hilbert term $M_{\rm pl}^2R/2$ 
in the Horndeski Lagrangian (\ref{Lho}).

Compared to the leading-order tensor spectrum 
${\cal P}_h^{\rm lead}(k)=h^2/(4\pi^2 q_{\rm t}c_{\rm t}^3)$ in 
the original frame, the leading-order spectrum 
$\hat{{\cal P}}_h^{\rm lead}(k)=2\hat{H}^2/(\pi^2 M_{\rm pl}^2)$ 
in the Einstein frame depends on the Hubble 
parameter $\hat{H}$ alone.
This means that, even for very general inflationary models 
in the framework of GLPV theories, 
there is a frame in which the energy scale of inflation 
is directly known from the measurement of primordial 
gravitational waves. The disformal transformation provides 
us with the physical understanding that the amplitude of primordial 
tensor perturbations is intrinsically related to the expansion 
rate of the Universe in the Einstein frame.

Note that the leading-order tensor power spectrum in the Einstein frame 
was also found in Ref.~\cite{Cre} for the specific case with $\Omega^2=1$. 
We have further shown that the tensor spectrum in the Einstein frame 
matches with that appearing in GR even up to next-to-leading order 
in slow-roll for the more general transformation with $\Omega^2 \neq 1$.

Since $\hat{\bar{N}}^2=\Omega^2c_{\rt k}^2=8q_{\rt k}c_{\rt k}^3/M_{\rm pl}^2$ 
for the choices (\ref{ck}) and (\ref{qk}), 
the scalar power spectrum (\ref{Pzeta}) in the Einstein frame reads
\be
\hat{\cal P}_{\zeta} (k)=\frac{q_{\rt k}c_{\rt k}^3}{q_{\rs k}c_{\rs k}^3} 
\frac{\hat{H}^2}{\pi^2 M_{\rm pl}^2}
\left[ 1-2(C+1) \hat{\epsilon}_H +2(C+1)\hat{s}_{\rt}
-C\hat{\epsilon}_{\rs}-(3C+2)\hat{s}_{\rs}
\right]|_{\hat{c}_{\rs}k=\hat{a}\hat{h}}\,,
\label{PzetaEin}
\ee
where we used Eqs.~(\ref{qstra}), (\ref{cstra}), and (\ref{hateph2}). 
$\hat{\cal P}_{\zeta} (k)$ depends not only on $\hat{H}$ but also 
on the ratio $q_{\rt k}c_{\rt k}^3/(q_{\rs k}c_{\rs k}^3)$.

If $c_{\rs}<c_{\rt}$, the tensor-to-scalar ratio (\ref{ratio})
in the Einstein frame evaluated at $\hat{c}_{\rt}k=\hat{a}\hat{h}$ is given by 
\be
\hat{r}=2 \frac{q_{\rs}c_{\rs}^3}{q_{\rt}c_{\rt}^3} 
\left[ 1-2(C+1) \hat{s}_{\rt}+C\hat{\epsilon}_{\rs}
+(3C+2) \hat{s}_{\rs}+
(2\hat{\epsilon}_H-2\hat{s}_{\rt}+\hat{\epsilon}_{\rs}
+3\hat{s}_{\rs}) \ln \frac{c_{\rs}}{c_{\rt}} 
\right] \biggr|_{k=\hat{a}\hat{H}}\,,
\label{PrEin}
\ee
where we employed Eqs.~(\ref{eprt}) and (\ref{hateph2}).

If $c_{\rs}>c_{\rt}$, the evaluation of $\hat{r}$ should be performed
at ${\hat{c}_{\rs}k=\hat{a}\hat{h}}$ [see Eq.~(\ref{ratio2})], such that 
\be
\hat{r}=2 \frac{q_{\rs}c_{\rs}^3}{q_{\rt}c_{\rt}^3} 
\left[ 1-2(C+1)\hat{s}_{\rt}+C\hat{\epsilon}_{\rs}
+(3C+2) \hat{s}_{\rs}+
2\hat{\epsilon}_H \ln \frac{c_{\rs}}{c_{\rt}} 
\right] \biggr|_{\hat{c}_{\rs}k=\hat{a}\hat{h}}\,.
\label{PrEin2}
\ee

\subsection{Background equations of motion in the Einstein frame and 
conditions for the red-tilted tensor power spectrum}
\label{Einred}

{}From Eq.~(\ref{Phein}) the spectral index of the leading-order 
tensor power spectrum $\hat{{\cal P}}^{\rm lead}_h(k)
=2\hat{H}^2/(\pi^2 M_{\rm pl}^2)$ in the Einstein frame
is simply given by 
\be
\hat{n}_{\rt}^{\rm lead}=-2\hat{\epsilon}_{H}\,.
\ee
The tensor spectrum is red-tilted ($\hat{n}_{\rt}^{\rm lead}<0$) 
under the condition 
\be
\hat{\epsilon}_{H}>0\,,\qquad {\rm i.e.,} \qquad
\dot{\hat{H}}<0\,.
\label{red}
\ee
The same condition was also derived in Ref.~\cite{Cre} without 
specifying gravitational theories.
In the following, we translate the condition $\dot{\hat{H}}<0$ explicitly 
in GLPV theories by considering the background equations of motion.

In the transformed frame, 
Eqs.~(\ref{back1}) and (\ref{back2}) read
\ba
\hspace{-0.3cm}
& & \hat{A}_2-6\hat{H}^2 \hat{A}_4-12\hat{H}^3 \hat{A}_5
+\hat{\bar{N}} \left( \hat{A}_{2,\hat{N}}+3\hat{H}\hat{A}_{3,{\hat{N}}}
+6\hat{H}^2 \hat{A}_{4,\hat{N}}+6\hat{H}^3 \hat{A}_{5,\hat{N}} \right)=0\,,
\label{back1d} \\
\hspace{-0.3cm}
& & \hat{A}_2-6\hat{H}^2 \hat{A}_4-12\hat{H}^3 \hat{A}_5
-\frac{1}{\hat{\bar{N}}} \left( \dot{\hat{A}}_3+4\dot{\hat{H}}\hat{A}_4 
+4\hat{H} \dot{\hat{A}}_4+12\hat{H} \dot{\hat{H}}\hat{A}_5
+6\hat{H}^2\dot{\hat{A}}_5 \right)=0\,.
\label{back2d}
\ea
On using $\hat{q}_{\rt}=\hat{L}_{,{\hat{\cal S}}}/(4\hat{\bar{N}})$, 
the choice (\ref{qk}) corresponds to 
$\hat{L}_{,{\hat{\cal S}}}=M_{\rm pl}^2/2$, i.e., 
\be
\hat{A}_4=-\frac{M_{\rm pl}^2}{2}
-3\hat{H} \hat{A}_5\,.
\label{hatA4d}
\ee
Since $\hat{c}_{\rt}^2=\hat{\bar N}^2 \hat{\cal E}/\hat{L}_{,\hat{\cal S}}$ 
and $\hat{{\cal E}}=\hat{B}_4+\dot{\hat{B}}_5/(2\hat{\bar N})$, the choice 
(\ref{ck}) gives rise to another relation for $\hat{B}_4$ and $\hat{B}_5$. 
However Eqs.~(\ref{back1d}) and (\ref{back2d}) 
do not contain $\hat{B}_4$, $\hat{B}_5$ and their derivatives, 
so the background equations of motion are not affected by choosing 
$\hat{c}_{\rt k}$ as Eq.~(\ref{ck}).
Substituting Eq.~(\ref{hatA4d}) into Eqs.~(\ref{back1d}) and (\ref{back2d}), 
the terms $3M_{\rm pl}^2\hat{H}^2$ and $2M_{\rm pl}^2 \dot{\hat{H}}$ arise 
from $-6\hat{H}^2 \hat{A}_4$ and $-4\dot{\hat{H}} \hat{A}_4$ respectively.
Then we can express Eqs.~(\ref{back1d}) and (\ref{back2d}) 
in the following forms
\ba
3M_{\rm pl}^2 \hat{H}^2 &=& \hat{\rho}\,,\label{backE1}\\
-2M_{\rm pl}^2 \frac{1}{\hat{\bar N}}\frac{d\hat{H}}{dt} &=&
\hat{\rho}+\hat{P}\,,\label{backE2}
\ea
where 
\ba
\hat{\rho} &\equiv& 
-\hat{A}_2-6\hat{H}^3 \hat{A}_5-
\hat{\bar{N}} \left( \hat{A}_{2,\hat{N}}+3\hat{H}\hat{A}_{3,{\hat{N}}}
-12\hat{H}^3 \hat{A}_{5,\hat{N}} \right)\,,\\
\hat{P} &\equiv&
\hat{A}_2+6\hat{H}^3\hat{A}_5
-\frac{1}{\hat{\bar N}}
\left( \dot{\hat{A}}_3-12\hat{H}\dot{\hat{H}} \hat{A}_5
-6\hat{H}^2\dot{\hat{A}}_5  \right)\,.
\ea
The background equations of motion (\ref{backE1}) and (\ref{backE2}) 
correspond to those appearing in Einstein gravity. 
{}From these equations the effective energy density $\hat{\rho}$ 
and the pressure $\hat{P}$ obey the continuity equation 
\be
\frac{1}{\hat{\bar N}}\frac{d}{dt} \hat{\rho}
+3\hat{H} \left( \hat{\rho}+\hat{P} \right)=0\,.
\ee
{}From Eq.~(\ref{backE2}) the condition (\ref{red}) translates 
to $\hat{\rho}+\hat{P}>0$, i.e., 
\be
\hat{\bar{N}} \left( \hat{A}_{2,\hat{N}}+3\hat{H}\hat{A}_{3,{\hat{N}}}
-12\hat{H}^3 \hat{A}_{5,\hat{N}} \right)
+\frac{1}{\hat{\bar N}}
\left( \dot{\hat{A}}_3-12\hat{H}\dot{\hat{H}} \hat{A}_5
-6\hat{H}^2\dot{\hat{A}}_5 \right)<0\,,
\label{redcon}
\ee
under which the tensor power spectrum is red-tilted. 
The explicit condition (\ref{redcon}) is useful to confront inflationary 
models in the framework of GLPV theories with the observations of CMB.

\section{Conclusions}
\label{consec}

On the flat FLRW background we have shown that the curvature 
perturbation $\zeta$ and the tensor perturbation $\gamma_{ij}$ are 
invariant under the disformal transformation 
$\hat{g}_{\mu \nu}=\Omega^2 (\phi) g_{\mu \nu}
+\Gamma (\phi,X) \nabla_{\mu} \phi 
\nabla_{\nu} \phi$
by choosing the unitary gauge $\phi=\phi(t)$.
This is the generalization of Ref.~\cite{Minami} in which the same 
property was also found for the transformation of the form 
$\hat{g}_{\mu \nu}=\Omega^2 (\phi) g_{\mu \nu}
+\Gamma (\phi) \nabla_{\mu} \phi 
\nabla_{\nu} \phi$. 
While the latter transformation preserves the structure of 
Horndeski theories, the former can deal with the transformation 
between more general theories beyond Horndeski, e.g.,
GLPV theories.

In unitary gauge the Lagrangian $L$ of GLPV theories on the flat FLRW 
background is given by Eq.~(\ref{LH}), which depends on the lapse 
$N$, the time $t$, and other three-dimensional geometric scalars 
$K,{\cal S},{\cal R},{\cal U}$. 
Under the disformal transformation (\ref{distra}), the structure of the action 
$S=\int d^4 x \sqrt{-g}L$ is preserved with the coefficients related 
to each other as Eqs.~(\ref{hatA2})-(\ref{hatB5}). 
The relations (\ref{Hore1}) and (\ref{Hore2}) imply that the transformation 
between Horndeski theories is required to satisfy the condition $\Gamma_{,X}=0$, 
i.e., $\Gamma=\Gamma(\phi)$.

Expanding the action (\ref{geneac}) up to quadratic order in scalar and tensor 
perturbations, we have derived the corresponding second-order 
Lagrangian densities (\ref{L2}) and (\ref{L2ten}), respectively. 
Unlike Refs.~\cite{Piazza,Gergely} we have explicitly taken into account the 
background value of the lapse $N$. 
This is important for studying the relations of physical quantities 
between the two metric frames connected
under the disformal transformation. 
In GLPV theories we have presented explicit relations between the two frames 
for the quantities associated with the background and perturbation 
equations of motion.
In particular, the quantities $q_{\rs},c_{\rs}^2,q_{\rt},c_{\rt}^2$, 
which are associated  with conditions for the absence of ghosts 
and Laplacian instabilities, transform as Eqs.~(\ref{qstra}), (\ref{cstra}), 
(\ref{qttra}), and (\ref{cttra}), respectively.

In Sec.~\ref{powerspssec} we have obtained the next-to-leading order 
inflationary power spectra of curvature and tensor perturbations 
as well as their spectral indices in the forms (\ref{Pzeta}), (\ref{Ph}), 
(\ref{ns}), and (\ref{nt}), respectively. 
For $c_{\rs}<c_{\rt}$ the tensor-to-scalar ratio $r$ is given by 
Eq.~(\ref{ratio}), whereas for $c_{\rs}>c_{\rt}$ it is expressed as 
Eq.~(\ref{ratio2}).
We have explicitly proved that the inflationary observables are 
invariant under the disformal transformation up to next-to-leading 
order in slow-roll.

In Sec.~\ref{einsteinsec} we have identified the existence of the Einstein 
frame in which the next-to-leading order tensor power spectrum is in
the same form as that in Einstein frame. 
The power spectra of tensor and scalar perturbations in this frame
are given, respectively, by Eqs.~(\ref{Phein}) and (\ref{PzetaEin}).
In the Einstein frame the function $\hat{A}_4$ in GLPV theories 
has the relation $\hat{A}_4=-M_{\rm pl}^2/2-3\hat{H}\hat{A}_5$, 
in which case the Lagrangian (\ref{Lho}) in Horndeski theories 
involves the Einstein-Hilbert term $M_{\rm pl}^2 R/2$.
In GLPV theories the condition under which the leading-order 
tensor power spectrum is red-tilted is characterized 
by Eq.~(\ref{redcon}).

Finally, we summarize the main results of our paper together with 
further possible applications.

\begin{itemize}

\item
The next-to-leading order scalar and tensor power spectra derived in 
this paper are useful to place tight and precise constraints on a wide 
variety of single-field inflationary models in the framework of 
GLPV theories (along the lines of Refs.~\cite{Kuro,Planck2}). 
In particular, the future possible detection of 
primordial gravitational waves will allow us to determine the 
inflationary Hubble parameter $\hat{H}$ in the Einstein frame
appearing in the tensor power spectrum (\ref{Phein}).

\item
We have shown the invariance of curvature and tensor perturbations 
under the disformal transformation in the single-field inflationary scenario.
In BD theories, it was further proved that other dimensionless cosmological 
observables--such as the redshift, the reciprocity relation, 
temperature anisotropies--are conformally independent of the chosen 
metric frames \cite{Chiba2,Catena}.
We expect that the similar properties for the invariance of 
dimensionless observables would also hold for GLPV theories 
under the disformal transformation, but the detailed study will be 
necessary to understand the correspondence between physical 
quantities in different metric frames.

\item
In the context of dark energy we need to take into account 
additional matter fields to the Lagrangian.
In such cases the propagation speeds of the scalar field $\phi$
and matter are mixed each other even for the metric frame 
minimally coupled to matter \cite{Gergely,Gleyzes,GlHa,Kase2}.
This non-trivial mixing can be understood by 
the disformal transformation to the Einstein frame
under which the factor $\Gamma$ involving the $X$ dependence
gives rise to a kinetic-type coupling of the scalar field 
with matter \cite{Garcia,Minami}.
It will be of interest to study the role of such a specific coupling 
and resulting observational consequences.

\end{itemize}

\begin{acknowledgments}
The author thanks Antonio De Felice for useful discussions.
The author is supported by the Grant-in-Aid for Scientific Research from 
JSPS (No.~24540286), the i-Link cooperation program of CSIC (project ID i-Link0484), 
and the cooperation programs between Tokyo University of Science and CSIC.
\end{acknowledgments}

\end{document}